\documentclass[letterpaper]{article}
\usepackage[T1]{fontenc}
\usepackage[margin=1in]{geometry}
\usepackage{graphicx}
\usepackage{xcolor}
\usepackage{subcaption}
\usepackage{flafter}
\usepackage[section]{placeins}
\usepackage{amsmath,amssymb}
\usepackage{siunitx}
\usepackage{authblk}
\usepackage[colorlinks=true,allcolors=blue]{hyperref}
\usepackage[style=chem-acs,doi=false,url=true,isbn=false]{biblatex}
\author[1]{X. Romain}
\author[1,2]{A. P. Vafa}
\author[1]{M. Florescu*}
\affil[1]{Optoelectronics Research Centre, University of Southampton, Southampton, UK}
\affil[2]{Advanced Technology Institute, University of Surrey, Guildford, UK}

\title{Spacer-Engineered Spectral Selectivity \\in Metal--Carbon Solar--Thermal Absorbers}
\date{*Email: m.florescu@soton.ac.uk}

\begin{document}

\maketitle

\begin{abstract}
Spectrally selective solar--thermal absorbers must capture sunlight while limiting emission at operating temperatures where the Planck spectrum overlaps the solar band. We connect this trade-off to a controlled optical mechanism in a metal--carbon metasurface. A radiative-limit calculation using AM1.5D illumination identifies how cutoff wavelength, infrared absorptivity and concentration together constrain equilibrium temperature, while the metasurface calculations reveal the additional role of a sharp spectral transition. We then separate the dielectric gap from the absorbing carbon layer in a gold--carbon architecture. Frequency-domain fields and material-resolved dissipation show that the spacer controls the overlap of a gap-localised mode with carbon, extending the high-absorptivity band and sharpening its infrared roll-off. A thickness sweep demonstrates that the spectrum with the strongest visible absorption need not give the highest radiative-limit equilibrium temperature. Polar-angle- and polarisation-resolved calculations demonstrate the angular robustness of the proposed plasmonic metasurface and assess the extent to which normal-incidence spectra approximate hemispherical emission. The calculated high-temperature values are benchmarks for the fixed-geometry, fixed-optical-constant model, not operating temperature predictions for the gold structure. The study establishes spacer-controlled loss partitioning as a route to shaping solar--thermal spectra and distinguishes optical selectivity from materials durability and useful heat output.
\end{abstract}

\section*{Keywords}

solar-thermal absorbers; spectral selectivity; metasurfaces; plasmonic resonances; radiative equilibrium; thermal emission.

\section{Introduction}

Solar-thermal conversion transforms incident radiation directly into heat and underpins technologies ranging from solar collectors to thermophotovoltaic, thermoelectric and thermionic energy conversion~\cite{tianReviewSolarCollectors2013,kalogirouSolarThermalCollectors2004,Schwede2010Photon-enhancedSystems}. Unlike a single-junction photovoltaic device, whose useful absorption window is fixed by the semiconductor bandgap, a thermal absorber can in principle draw energy from the full solar spectrum. Its performance is therefore governed not by carrier generation but by the spectral and angular channels through which it exchanges radiation with its surroundings: the same reciprocal optical states that admit sunlight also determine the emissivity of the heated structure. This perspective has motivated multilayer coatings, metamaterials and metasurfaces that use interference, impedance matching, resonant confinement and waveguide coupling to approach unity absorptance~\cite{hanBroadbandSolarThermal2016,abed2017,rheeMetamaterialbasedPerfectAbsorbers2014,tomescuNanopillarsMetasurfaceModelled2020}. Related photonic-crystal and correlated-disorder strategies have likewise shown that absorption can be increased by engineering the optical mode spectrum and the momentum-space channels through which incident light couples into an absorber~\cite{florescuImprovingSolarCell2007,tavakoliOver65Sunlight2022,vafaBalancingOrderDisorder2026,lambertz2026}.

At moderate temperatures, broadband absorption may be sufficient. As the temperature rises, however, the Planck blackbody spectrum shifts towards the near infrared and increasingly overlaps the solar-absorbing band. Enhancing absorptance without regard to wavelength can then be counterproductive, because the same optical channels contribute to thermal emission~\cite{zhangSolarSelectiveAbsorber2022,fanThermalPhotonicsEnergy2017,liNanophotonicControlThermal2018}. High-temperature performance is consequently governed by the full thermally weighted spectrum: strong visible and near-infrared absorption must be combined with suppressed infrared emissivity, and both the position and the sharpness of the transition between these regimes become central design variables. Studies of finite photonic crystals have further established that thermal emission depends not only on local material dissipation but also on the available optical states, propagation channels and surface termination~\cite{schulerThermalEmissionFinite2009}. These considerations become particularly stringent above 1000~K and at moderate concentration factors, where even weak residual infrared absorptivity can produce substantial radiative loss.

Carbon films, ultrathin metals, nanowire networks, graphene structures and metal--dielectric--metal (MDM) metasurfaces have all been explored as routes to solar-selective absorption~\cite{tinchevThermalSolarAbsorber2010,ahmadUltrathinMetalFilms2012,ghobadi97PercentLight2017,linStructuredGrapheneMetamaterial2020,wangPerfectSelectiveMetamaterial2013a,wangHighlyEfficientSelective2015,xuNearperfectSpectrallyselectiveMetasurface2022}. MDM absorbers are particularly attractive because the nanoscale region between a patterned metal layer and a reflector can support strongly confined electromagnetic modes. In the closest metal--carbon--metal architectures, however, the amorphous-carbon interlayer simultaneously fixes the resonator separation, defines the local dielectric environment and provides a substantial fraction of the optical loss~\cite{wanBroadbandMetasurfaceAbsorber2015,wanSelectiveMetasurfaceAbsorber2016}. Modal confinement and material dissipation are therefore tied to the same physical layer, restricting the freedom with which the spectral transition can be shaped.

Dielectric gaps are already established in selective metal--dielectric--metal metasurfaces, including gold-based solar absorbers~\cite{azadMetasurfaceBroadbandSolar2016}, and are used to tune cavity and plasmonic coupling in other optical devices~\cite{Mao2020ManipulatingReflection,pillaiEffectDielectricSpacer2011,lalUsingSpacerLayers2012}. The distinction pursued here is the separation of two functions that coincide in a metal--carbon--metal absorber: the dielectric spacer controls the gap field, while a distinct carbon layer supplies much of the loss~\cite{wanBroadbandMetasurfaceAbsorber2015,wanSelectiveMetasurfaceAbsorber2016}. We test whether that separation alters \emph{where} the resonant energy is dissipated, not merely the position of an absorption peak. The comparison among a planar stack, a single resonator and the complete metasurface isolates the mechanism before its thermal consequences are evaluated.

We first use a radiative-limit balance to identify the spectral trade-off, then resolve the gap fields and loss in the separate Au and carbon components, and finally compare the resulting spectra after angular weighting. FDTD transients provide further insight into the dynamics of the energy transport (Supplementary Information), and a phenomenological figure of merit summarises the spacer-thickness trends. Gold provides a well-characterised optical model for this comparison; calculated equilibrium temperatures do not establish the thermal stability or useful heat output of an Au device.

\section{Radiative framework for high-temperature spectral selectivity}

At elevated temperature, spectral selectivity is constrained by Kirchhoff's law: any reciprocal optical channel that raises absorptance at a given wavelength and direction can also raise emissivity in the corresponding thermal channel. The design problem is therefore governed by the combined dependence on cutoff wavelength, operating temperature, concentration factor and residual infrared absorptivity, rather than by absorptance considered in isolation. We first connect local electromagnetic dissipation to the macroscopic energy balance and then use idealised cutoff spectra to map these dependencies across the relevant spectral window. This construction provides both a general framework for high-temperature selective absorbers and the reference against which the metasurface response is evaluated.

\subsection{From local optical dissipation to radiative equilibrium}
\label{sec:ch6:thermal-balance-theory}

\begin{figure}[!htbp]
    \centering
    \begin{subfigure}{0.42\linewidth}
        \centering
        \includegraphics[width=\linewidth]{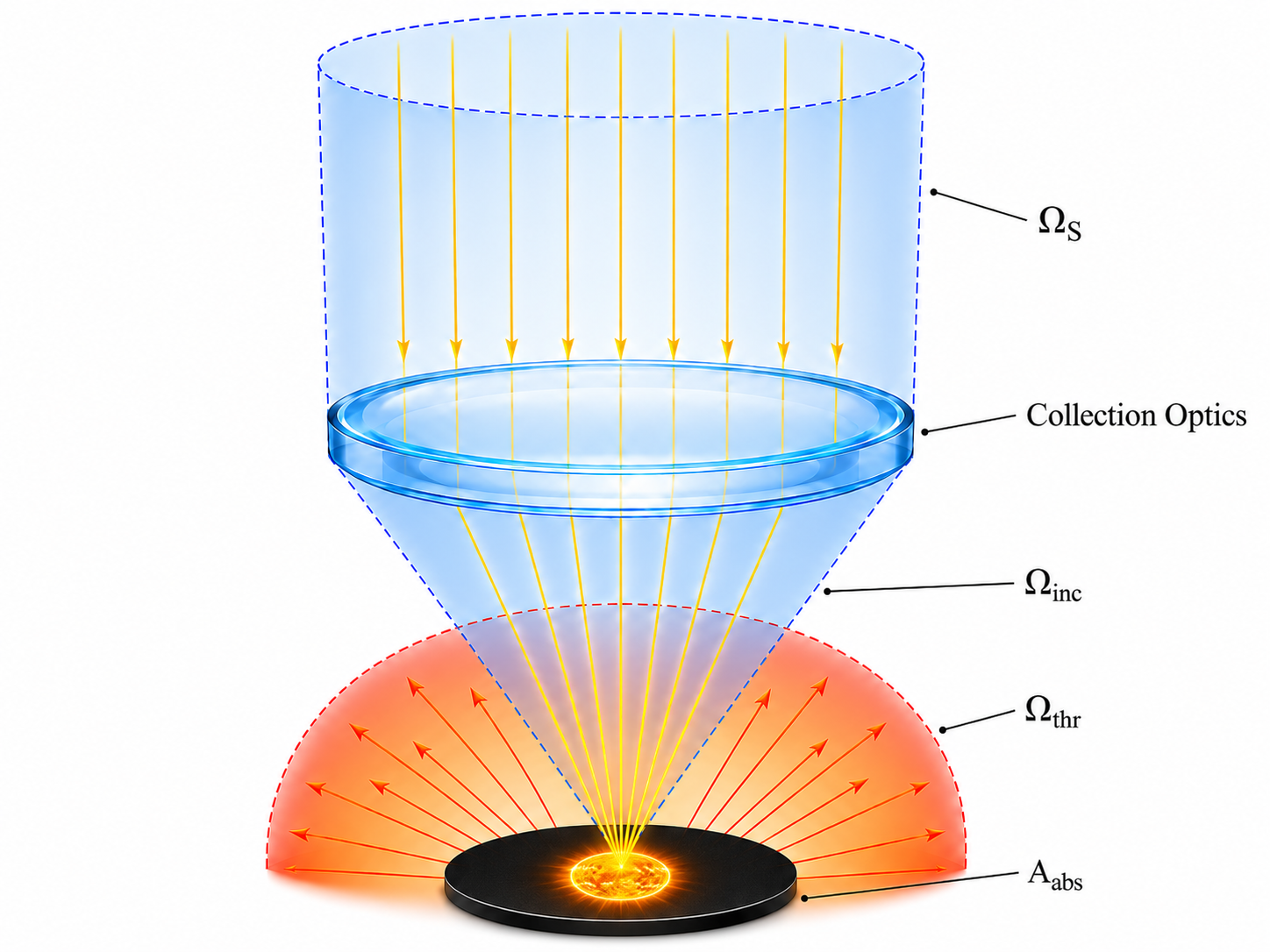}
        \caption{}
        \label{fig:solar-thermal-balance-schematic}
    \end{subfigure}
    \hfill
    \begin{subfigure}{0.56\linewidth}
        \centering
        \includegraphics[width=0.92\linewidth]{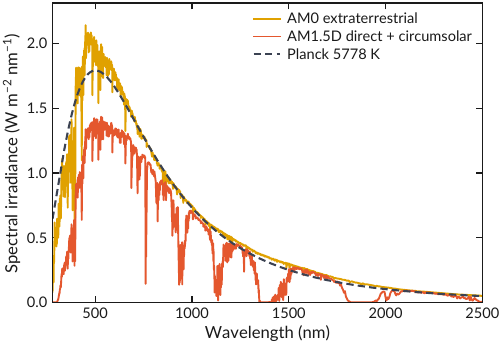}
        \caption{}
        \label{fig:ch6:solar-spectrum-am0-am15d-bb}
    \end{subfigure}
    \caption{(a) Radiative-balance model for a concentrated solar-thermal absorber. Solar radiation within $\Omega_\mathrm{s}$ is redirected into $\Omega_\mathrm{inc}$ and focused onto the projected area $A_\mathrm{abs}$, while thermal radiation leaves through $\Omega_\mathrm{thr}$. (b) AM0,  AM1.5D  and the 5778 K irradiance spectra used in the model.}
    \label{fig:schematic+solar}
\end{figure}

To connect the optical response of the metasurface with the temperature that it can sustain, we consider the concentrated solar-thermal configuration sketched in Fig.~\ref{fig:solar-thermal-balance-schematic}. At the Earth's surface, the Sun subtends the small solid angle
\(\Omega_\mathrm{s}\simeq\pi R_{\odot}^{2}/D_{SE}^{2}
=6.8\times10^{-5}\,\si{\steradian}\), where \(R_{\odot}\) is the solar radius and \(D_{SE}\) is the Sun--Earth distance. Collection optics redirect this radiation into an incident cone \(\Omega_\mathrm{inc}\) and focus it onto the projected absorber area \(A_\mathrm{abs}\). Once heated, the absorber exchanges energy with its surroundings through thermal emission into the outgoing solid angle \(\Omega_\mathrm{thr}\), together with any conductive, convective, or useful heat-extraction channels. The incident spectrum used throughout is the AM1.5D direct-plus-circumsolar irradiance \(I_{\lambda\odot}\), shown in Fig.~\ref{fig:ch6:solar-spectrum-am0-am15d-bb}. This spectrum is appropriate for concentrated illumination because it contains the direct and circumsolar radiation accessible to the collection optics, whereas AM1.5G also includes the diffuse sky component~\cite{Steiner2016RatingRobins}.

Solar-thermal conversion proceeds by dissipating optical energy into electronic and lattice degrees of freedom. This contrasts with photovoltaic operation, where absorbed photons are converted into extractable carriers and subsequent thermalisation constitutes a loss. Here, non-radiative electronic relaxation, electron--phonon coupling, lattice heating and Ohmic dissipation are the desired microscopic pathways~\cite{baffouThermoplasmonicsMetallicNanostructures2013,incroperaFundamentalsHeatMass2017}. For a non-magnetic lossy medium, the time-averaged power dissipated per unit volume is
\begin{equation}
    Q(\mathbf{r},\omega)
    =
    \frac{\omega}{2}\epsilon_0\epsilon''(\mathbf{r},\omega)
    \left|\mathbf{E}(\mathbf{r},\omega)\right|^2,
    \label{eq:absorbed-power-density}
\end{equation}
where $\omega$ is the angular frequency, $\epsilon_0$ is the vacuum permittivity, $\epsilon''$ is the imaginary part of the relative permittivity and $\mathbf{E}$ is the local electric field. Integrating $Q$ over a material volume provides the power dissipated in that region. For a plane wave of wavelength $\lambda$, propagation direction $\mathbf{n}(\theta,\phi)$ and polarisation $\mathbf{p}$, the directional absorptivity $\alpha(\lambda,\mathbf{n},\mathbf{p})$ is the absorbed power normalised to the incident power. Equation~\ref{eq:absorbed-power-density} therefore provides the microscopic link between field localisation and heat generation. We assume that the ultrathin stack thermalises rapidly compared with the radiative dynamics and can be represented by a spatially uniform temperature.

At the macroscopic level, the internal energy $U$ obeys
\begin{equation}
    \frac{dU}{dt}
    =
    P_\mathrm{abs}
    -P_\mathrm{rad}
    -P_\mathrm{cond}
    -P_\mathrm{conv}
    -P_\mathrm{useful},
    \label{eq:general-energy-balance}
\end{equation}
where the terms on the right denote absorbed optical power, radiative loss, conductive loss, convective loss and useful extracted heat, respectively~\cite{incroperaFundamentalsHeatMass2017,modestRadiativeHeatTransfer2021}. Equations~\ref{eq:absorbed-power-density} and~\ref{eq:general-energy-balance} define the two levels of description used throughout: the first resolves where energy is dissipated inside the nanostructure, whereas the second tracks how the integrated absorbed power is partitioned at steady state. Here, we adopt the idealised radiative limit, setting $P_\mathrm{cond}=P_\mathrm{conv}=P_\mathrm{useful}=0$. The resulting temperatures are radiative-limit equilibrium values for a fixed optical structure; they do not include parasitic heat loss, temperature-dependent optical constants or thermally induced changes in geometry. The absorbed solar power is
\begin{equation}
    P_\mathrm{abs}
    = CA_\mathrm{abs}\int_0^\infty
    \alpha(\lambda)I_{\lambda\odot}(\lambda)\,d\lambda,
    \label{eq:absorbed-solar-power}
\end{equation}
where $C = A_a/A_{abs}$ or $C=\Omega_\mathrm{inc}/\Omega_\mathrm{s}$ is the geometric concentration factor for ideal imaging, $A_a$ the collector's aperture area, $A_\mathrm{abs}$ is the projected absorber area, $I_{\lambda\odot}$ is the unconcentrated AM1.5D spectral irradiance, unless noted otherwise, and $\alpha(\lambda)$ is the absorptivity used in the scalar reference calculation. Optical losses in the concentrator are represented by an efficiency $\eta_\mathrm{opt}\in[0,1]$, defined as the fraction of collected solar power delivered to the absorber; equivalently, $C$ is replaced by $\eta_\mathrm{opt}C$. Unless stated otherwise, we assume $\eta_\mathrm{opt}=1$.

The blackbody spectral radiance per unit solid angle, projected area and wavelength is
\begin{equation}
    B_\lambda(\lambda,T)
    =
    \frac{2hc^2}{\lambda^5}
    \frac{1}{\exp\left(hc/\lambda k_\mathrm{B}T\right)-1},
    \label{eq:planck-radiance}
\end{equation}
and the radiative power emitted at temperature $T_w$ is obtained by integrating over wavelength and the outgoing solid angle,
\begin{equation}
    P_\mathrm{rad}(T_w)
    =
    A_\mathrm{abs}\int_0^\infty
    \int_{\Omega_\mathrm{thr}}
    \overline{\varepsilon}(\lambda,\mathbf{n})
    B_\lambda(\lambda,T_w)
    \cos\theta\,d\Omega\,d\lambda,
    \label{eq:angular-radiative-loss}
\end{equation}
where $A_\mathrm{abs}$ is also the projected emitting area, $\mathbf{n}(\theta,\phi)$ specifies the emission direction and $\overline{\varepsilon}$ is the polarisation-averaged directional emissivity,
\begin{equation}
    \overline{\varepsilon}(\lambda,\mathbf{n})
    =
    \frac{
    \varepsilon_P(\lambda,\mathbf{n})
    +
    \varepsilon_S(\lambda,\mathbf{n})
    }{2}.
    \label{eq:polarisation-averaged-emissivity}
\end{equation}
Here $\varepsilon_P$ and $\varepsilon_S$ denote emissivities in the two orthogonal linear polarisations defined relative to the plane of incidence. The factor of two in Planck's law already accounts for both electromagnetic polarisations, so their directional emissivities enter through the average rather than the sum. 

For a passive reciprocal structure in local thermal equilibrium, Kirchhoff's law equates the emissivity of an outgoing channel to the absorptivity of its time-reversed incoming channel. Labelling reciprocal incoming and outgoing channels by the same surface-relative angles, for the planar geometry considered here,
\begin{equation}
    \varepsilon_{P,S}(\lambda,\theta,\phi)
    =
    \alpha_{P,S}(\lambda,\theta,\phi),
    \label{eq:kirchhoff-law}
\end{equation}
where $\alpha_P$ and $\alpha_S$ are the directional absorptivities for the corresponding polarisations. Thus, resonances that improve solar absorption can also increase thermal emission when they overlap the wavelengths populated by the hot absorber~\cite{modestRadiativeHeatTransfer2021,liNanophotonicControlThermal2018}. In finite periodic structures, this emission is further shaped by the available propagating states and by the termination of the photonic medium~\cite{schulerThermalEmissionFinite2009}. 

For the initial scalar reference calculation, we approximate the response as direction independent and integrate over the outward hemisphere $\Omega_\mathrm{thr}=2\pi~\mathrm{sr}$,
\begin{equation}
    P_\mathrm{rad}(T_w)
    =
    \pi A_\mathrm{abs}\int_0^\infty
    \alpha(\lambda)B_\lambda(\lambda,T_w)\,d\lambda.
    \label{eq:hemispherical-radiative-loss}
\end{equation}
Combining Eqs.~\ref{eq:general-energy-balance}, \ref{eq:absorbed-solar-power} and~\ref{eq:hemispherical-radiative-loss}, and including the ambient radiative background at $T_0=\SI{300}{\kelvin}$, gives
\begin{equation}
    C\int_0^\infty
    \alpha(\lambda)I_{\lambda\odot}(\lambda)\,d\lambda
    =
    \pi\int_0^\infty
    \alpha(\lambda)\left[B_\lambda(\lambda,T_w)-B_\lambda(\lambda,T_0)\right]d\lambda.
    \label{eq:radiative-limit-balance}
\end{equation}
This balance makes the design constraint explicit: increasing absorptivity is beneficial only where the additional solar power outweighs the corresponding increase in thermal emission. The optimal spectrum must therefore combine strong absorption across the solar band with suppressed emissivity in the longer-wavelength region that dominates radiative loss~\cite{fanThermalPhotonicsEnergy2017,liNanophotonicControlThermal2018}. The ambient term is retained throughout, although it becomes small in the high-temperature regime.

\subsection{Spectral design criteria in the radiative limit}
\label{Sec:aspect}

An ideal step spectrum, $\alpha=1$ below $\lambda_c$ and zero above it, illustrates the competition between extra solar capture and thermal emission (Fig.~S5). Extending the cutoff beyond roughly 2~$\mu$m adds little AM1.5D absorption while increasing thermal loss. The radiative balance, rather than a cutoff chosen from the solar spectrum alone, determines the useful transition.

\begin{figure}[!htbp]
    \centering
    \includegraphics[width=1\linewidth]{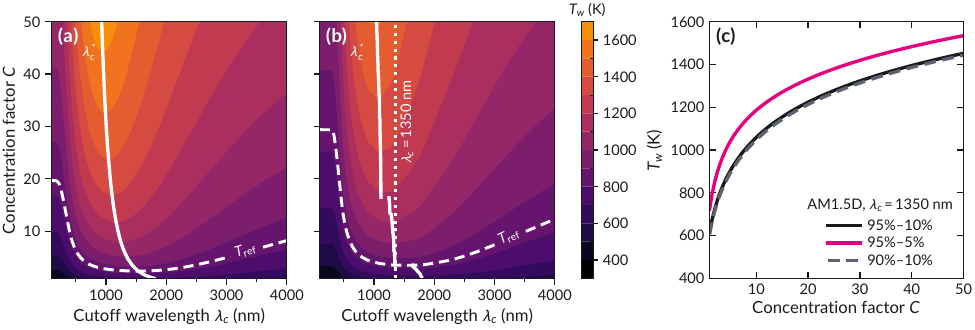}
    \caption{Radiative-limit equilibrium temperature $T_w$ of the two-level reference absorber as a function of cutoff wavelength and concentration factor for $\alpha_{\mathrm{up}}=0.95$ and $\alpha_{\mathrm{low}}=0.10$, (a) for a 5778~K blackbody sun diluted by the solar solid angle and (b) for the AM1.5D irradiance, in ambient temperature $T_0=0$ and $300$~K, respectively, on the same colour scale. The white dashed contour marks $T_{\mathrm{ref}}=830$~K, defined later as the reference temperature of the FOM calibration; the white dotted line in (b) marks the chosen cutoff $\lambda_c=1350$~nm and the solid white line the optimal cutoff $\lambda_c^*(C)$ that maximises $T_w$ at each $C$. (c) Equilibrium temperature with AM1.5D at $\lambda_c=1350$~nm for three two-level profiles.}
    \label{fig:target-absorber}
\end{figure}

To convert this qualitative design window into a quantitative reference, we replace the perfect step function by a two-level profile,
\begin{equation}
    \alpha_t(\lambda) =
\begin{cases}
\  \alpha_{\mathrm{up}}, & \lambda \leq \lambda_c, \\
\  \alpha_{\mathrm{low}}, & \lambda > \lambda_c,
\end{cases}
\label{eq:target-profile}
\end{equation}
where $\alpha_{\mathrm{up}}$ and $\alpha_{\mathrm{low}}$ are the chosen plateau values (0.95 and 0.10 for the primary reference; 0.05 for the lower infrared plateau in the comparison). For each pair $(\lambda_c,C)$, we insert $\alpha_t(\lambda)$ into Eq.~\ref{eq:radiative-limit-balance} and solve implicitly for $T_w$ using the 5778 K blackbody model and AM1.5D measured irradiance.
The resulting maps in Fig.~\ref{fig:target-absorber}(a) and (b) show a maximum, at given $C$, whose cutoff $\lambda_c^*$ decreases with concentration (solid white lines). While the theoretical model varies smoothly, in the practical map, the discrete jump due to the atmospheric water band, where the AM1.5D irradiance vanishes, signifies a target cutoff near 1350~nm at moderate concentration $C\approx10-20$.
A section through this map at $\lambda_c=1350$~nm (Fig.~\ref{fig:target-absorber}(c)) shows that reducing the infrared absorptivity from 0.10 to 0.05 nearly halves the concentration required to reach a given temperature up to about 1000~K (factor 0.51--0.53), and lowers it by about 41\% at 1200~K. Suppressing residual infrared emissivity is therefore a substantially stronger lever on high-temperature performance than solar band absorption enhancement.

For the chosen AM1.5D spectrum, $T_0=300$~K and the radiative-limit assumptions above, the 95\%--10\% reference profile is optimised by a cutoff close to 1350~nm. This value is a benchmark for the present analysis rather than a universal optimum: realistic spectra retain a finite transition width, resonant structure and a wavelength-dependent infrared tail, all of which enter the radiative balance.

\section{Results and discussion}

\begin{figure}[!htbp]
    \centering
        \includegraphics[width=1.0\linewidth]{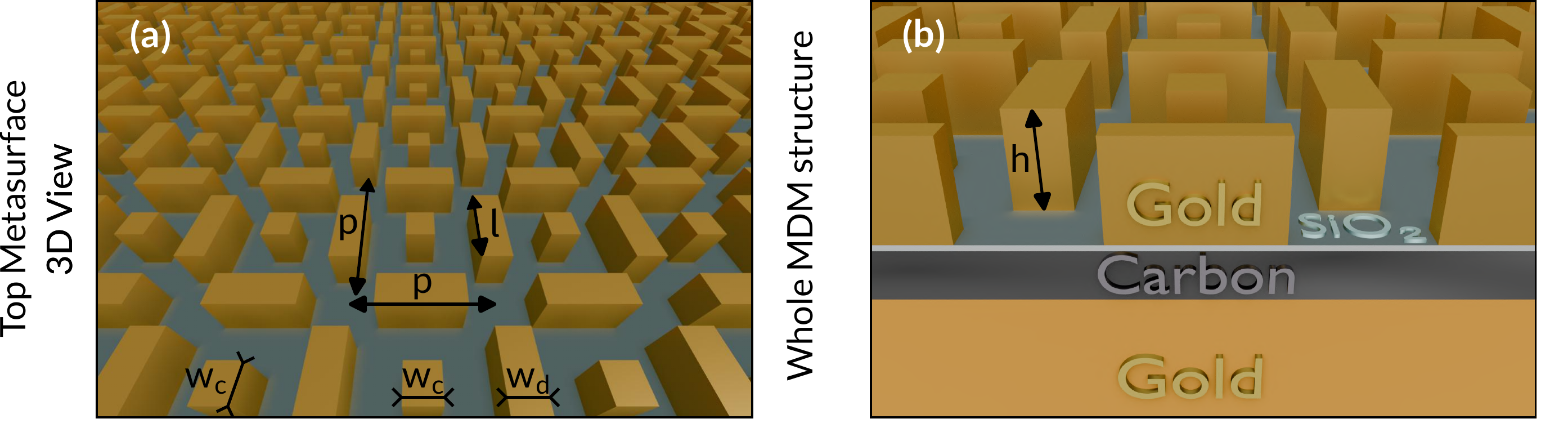}
    \caption{Spacer-engineered metal--carbon metasurface. (a) Patterned gold resonator layer with cuboid width $w_\mathrm{c}$, dipole length $l$ and dipole width $w_\mathrm{d}$. (b) Layer sequence comprising the gold resonators, idealised lossless dielectric spacer ($n=1.45$), absorbing carbon film and gold reflector.}
    \label{fig:Full-unitcell}
\end{figure}

The proposed architecture extends the established gold--carbon--gold metasurface~\cite{wanBroadbandMetasurfaceAbsorber2015,wanSelectiveMetasurfaceAbsorber2016} by inserting a transparent dielectric layer between the patterned resonators and the carbon absorber [Fig.~\ref{fig:Full-unitcell}(a,b)]. The square unit cell has period $p=\SI{200}{\nano\meter}$ and contains a central gold cuboid of lateral width $w_\mathrm{c}=\SI{50}{\nano\meter}$ together with two orthogonal dipoles of length $l=\SI{110}{\nano\meter}$ and width $w_\mathrm{d}=\SI{40}{\nano\meter}$. All resonators have height $h=\SI{70}{\nano\meter}$. Beneath the patterned layer lie a variable-thickness dielectric spacer, a \SI{35}{\nano\meter} amorphous-carbon film and a \SI{200}{\nano\meter} gold reflector. The spacer is modelled as an idealised lossless dielectric with real refractive index 1.45, representative of a SiO$_2$-like optical gap rather than a dispersive model of physical SiO$_2$; it can therefore alter the optical gap and modal field distribution without adding a comparable dissipative channel. The SiO$_2$ label in Fig.~\ref{fig:Full-unitcell}(b) denotes this idealised layer.

The gold and carbon optical constants are taken from Refs.~\cite{babarWeaver2015,hagemannOpticalConstantsFar1975} and are treated as temperature-independent; the corresponding refractive-index data are shown in Fig.~S1. Together with the ideal periodic geometry, this material model provides a controlled setting in which to isolate the electromagnetic role of the spacer. The precise numerical spectra will depend on the chosen material datasets and on fabrication-specific details, but the analysis is directed at the underlying coupling mechanism rather than at process optimisation. Surface roughness, sidewall taper and layer-thickness nonuniformity are outside the present scope. The model also excludes temperature-dependent optical constants, diffusion and morphological evolution. Bulk Au melts at 1337.33~K~\cite{nistGoldPoint}; therefore results approaching or exceeding this temperature are spectral and radiative-limit benchmarks, not feasible steady operating points for the present Au geometry. Stability below that limit is not established either.

\subsection{Isolating the spacer-mediated coupling mechanism}
\begin{figure}[!htbp]
    \centering
    \includegraphics[width=1\linewidth]{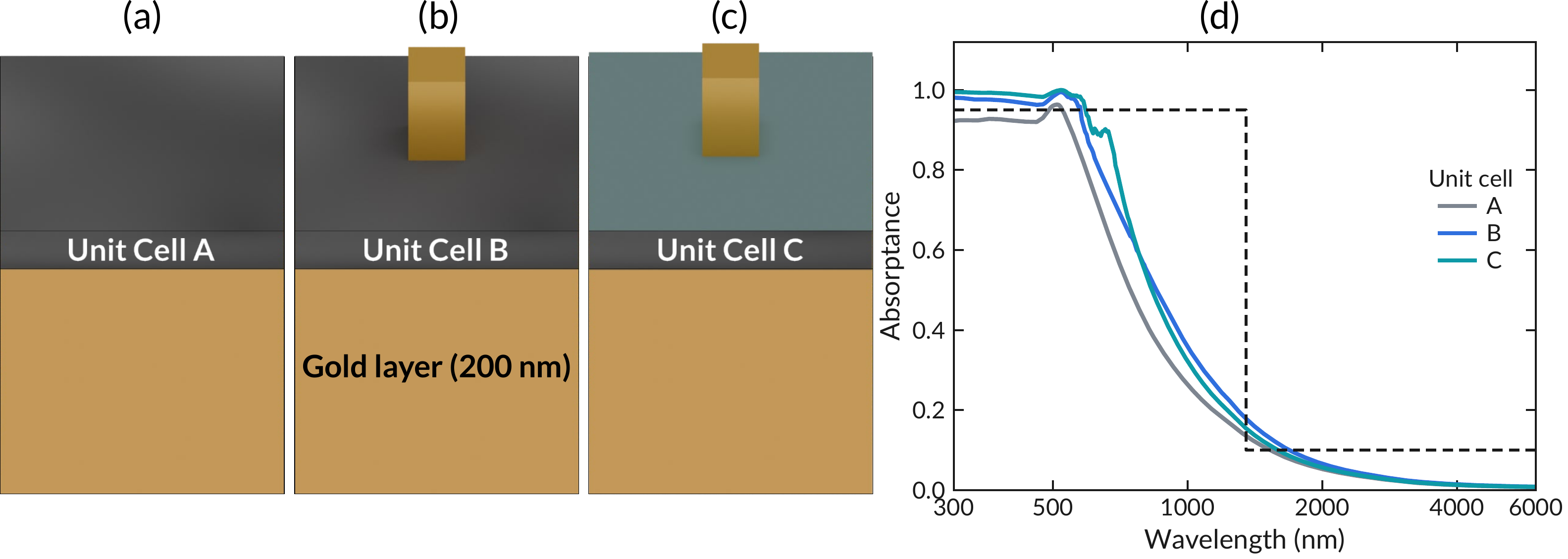}
    \caption{Intermediate structures used to isolate the spacer-mediated response: (a) carbon film on a gold reflector (unit cell A); (b) addition of a gold cuboid resonator (unit cell B); and (c) insertion of a 10 nm thick idealised lossless dielectric spacer between the resonator and carbon film (unit cell C). (d) Corresponding absorptance spectra.}
    \label{fig:intermediate-unitcells}
\end{figure}

All frequency-domain calculations are performed with the Wave Optics Module of COMSOL Multiphysics. To identify the role of the spacer before introducing interactions among several resonators of the full design, illustrated in Fig.~\ref{fig:Full-unitcell}, we construct a minimal hierarchy of intermediate structures: a planar carbon-on-gold stack (unit cell A), the same stack with a single gold cuboid (unit cell B), and the spacer-separated variant (unit cell C), as shown in Fig.~\ref{fig:intermediate-unitcells}. This progression isolates the change produced by the resonator and then by the transparent field-control layer. The corresponding spectra in Fig.~\ref{fig:intermediate-unitcells}(d) show that adding the cuboid in unit cell B increases absorptance across the visible and near-infrared, while introducing the spacer in unit cell C produces a further enhancement and sharpens the transition into the low-emissivity infrared regime. We verify numerical convergence for the proposed design (Supplementary Information; Section~SI.9); three meshes, with sharp and 5~nm-fillet-rounded corners, give nearly the same absorptance spectra, and their AM1.5D-weighted means in the resonant part of the spectrum at 450-1000~nm agree to within 0.01. The trends across the spacer sweep discussed below are the same on every mesh and for either corner shape. As an independent consistency check, the total absorptance obtained from $1-R-T$ was also compared with the volume-integrated dissipated power resolved over the constituent materials, confirming energy conservation within the numerical accuracy of the simulations ($4\times10^{-5}$ in absorptance over all angles, $1\times10^{-5}$ at normal incidence).
The spectral evolution across this hierarchy identifies the origin of the additional absorption. Relative to the planar reference, the cuboid in unit cell B broadens the visible and near-infrared response and introduces a resonance near $\lambda_0=\SI{600}{\nano\meter}$. A material-resolved decomposition (Fig.~S2) shows that the cuboid itself takes up a large part of the increase: averaged over the AM1.5D spectrum between 0.3 and 1.5~$\mu$m, the gold absorptance rises by 0.057 and the carbon absorptance by 0.030, the carbon gain being largest above the resonance (up to 0.13 near 760~nm). Inserting a \SI{10}{\nano\meter} spacer shifts the carbon absorption maximum to 675~nm and raises the mean absorptance between 300 and 500~nm from 0.97 to 0.99. At the same time, the long-wavelength roll-off becomes steeper without introducing an additional infrared background.
The single-cuboid field maps and line profiles at 633~nm (Fig.~S6) show that adding the spacer redistributes the gap field beneath the resonator. Together with the material-resolved spectra (Fig.~S2), this links the changed spectrum to field overlap with carbon rather than absorption in the nominally lossless spacer.

\subsection{Collective response of the complete metasurface}
\begin{figure}[!htbp]
    \centering
    \includegraphics[width=1\linewidth]{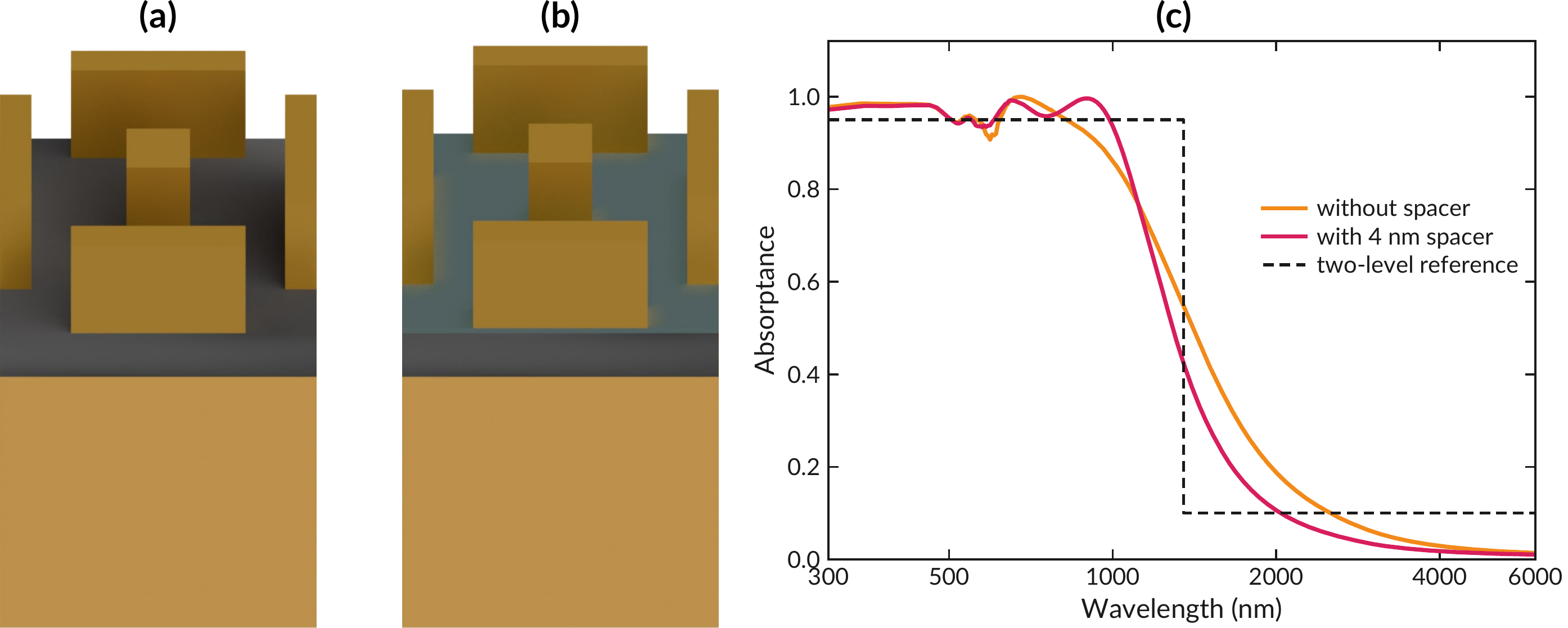}
    \caption{Complete metasurface unit cell (a) without and (b) with the idealised lossless dielectric spacer. (c) Corresponding absorptance spectra compared with the two-level reference profile.}%
    \label{fig:complete-unitcell-schematics+abs}
\end{figure}

The single-resonator hierarchy establishes the local-field-enhanced coupling mechanism. We now study the complete $C_4$-symmetric unit cell, shown in Fig.~\ref{fig:complete-unitcell-schematics+abs}(a),(b), to determine whether it survives interactions among the three resonant elements and can shape a broadband response.
The absorptance spectra in Fig.~\ref{fig:complete-unitcell-schematics+abs}(c) show that adding the \SI{4}{\nano\meter} spacer leaves the short-wavelength plateau essentially unchanged, moves the long-wavelength 95\% edge from about 820 to about 990~nm and steepens the decline beyond 1~$\mu$m. Thus, the spacer-controlled local mode survives in the coupled multi-resonator cell and reorganises the collective absorption band rather than producing an isolated narrow resonance.

\begin{figure}[!htbp]
    \centering
    \includegraphics[width=1\linewidth]{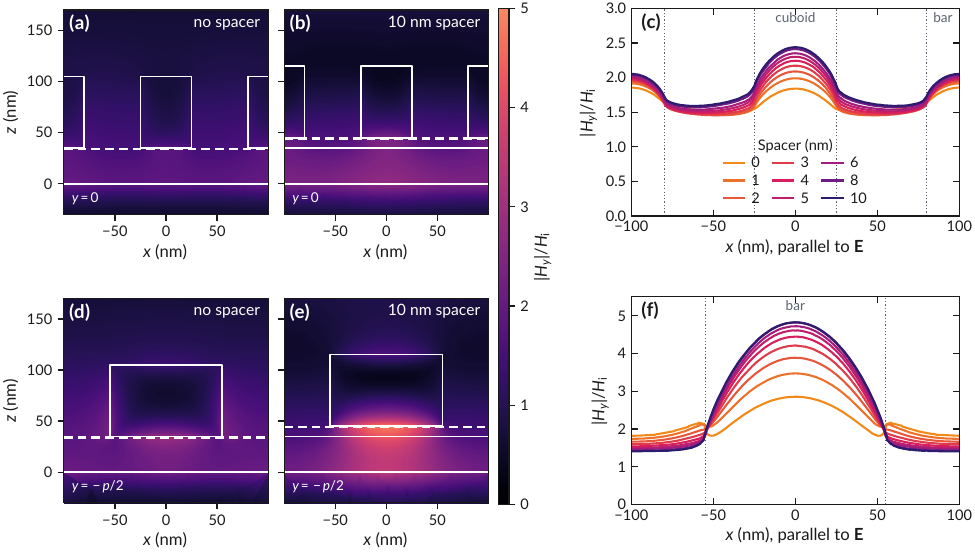}
    \caption{Magnetic field $|H_y|/H_\mathrm{i}$ in the $x$--$z$ plane through the centre of the unit cell ($y=0$) for the complete unit cell (a) without and (b) with the lossless dielectric spacer at $\lambda=987$~nm, and (d), (e) in the $x$--$z$ plane at the cell edge ($y=-p/2$) through the $x$-oriented resonator. (c) Corresponding $H_y$ line profiles beneath the resonator for increasing spacer thickness, through the cell centre beneath the cuboid, and (f) at $y=-p/2$ beneath the $x$-oriented resonator, along $x$ (parallel to the incident electric field) 1~nm below the resonator base.  All fields are normalised by the amplitude of the normally incident magnetic field, $H_\mathrm{i}$.}
    \label{fig:Fig7}
\end{figure}

Figure~\ref{fig:Fig7}(a) and (b) present magnetic field profile $|H_y|$, normalised by the normally incident amplitude $H_\mathrm{i}$, in the $x$--$z$ plane through the centre of the unit cell ($y=0$), and Fig.~\ref{fig:Fig7}(d) and (e) in the $x$--$z$ plane at the cell edge ($y=-p/2$), without and with 10~nm spacer, respectively.
The line profiles along the incident electric field, 1~nm below the resonator base, through the cell centre [Fig.~\ref{fig:Fig7}(c)] and at the cell edge [Fig.~\ref{fig:Fig7}(f)], reveal the same field reorganisation found in the minimal geometry, where the field is strongly localised below the resonator when the dielectric spacer layer is included. The spacer transforms the diffuse plasmon-like distribution of the no-spacer structure into a more strongly confined gap field, and the line profiles evolve continuously with thickness rather than switching discontinuously between distinct states. Beneath the cuboid and the $x$-oriented resonator, $|H_y|$ peaks at their centres and forms a standing wave. This progression supports the interpretation that the spacer tunes the overlap of an existing resonant mode with the carbon layer.

\begin{figure}[!htbp]
    \centering
    \includegraphics[width=0.85\linewidth]{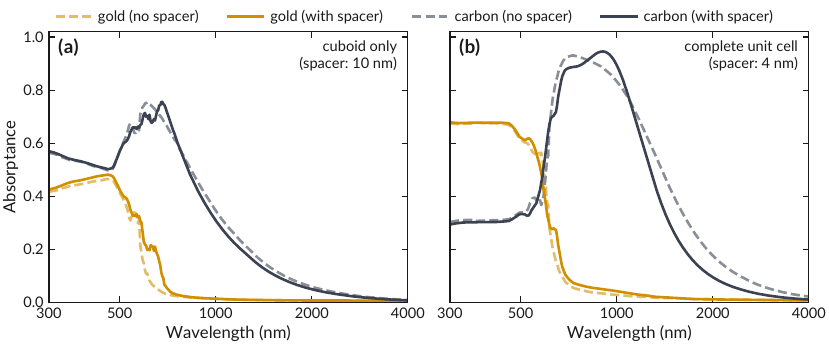}
    \caption{Material-resolved absorptance in the gold and carbon regions for (a) the cuboid-only structure without and with a 10~nm spacer and (b) the complete unit cell without and with a 4~nm spacer.}%
    \label{fig:NewFig8}
\end{figure}

The material-resolved spectra in Fig.~\ref{fig:NewFig8} resolve how this field redistribution changes dissipation in the gold and carbon regions. For the cuboid-only structure with a 10~nm spacer, sharper features appear in both the gold and carbon contributions. In the complete unit cell with a 4~nm spacer, by contrast, the dominant change occurs in the carbon spectrum, whose maximum moves from 730 to 905~nm and whose decline beyond 1.2~$\mu$m steepens, whereas the gold absorptance is altered only weakly. This redistribution is consistent with enhanced dissipation in the carbon layer rather than a comparable increase in metallic Ohmic loss. The spacer therefore changes where the resonant energy is dissipated: it drives a larger fraction of the gap-localised field into the absorbing carbon and sharpens the spectral transition.

Time-domain FDTD calculations resolve the route of the absorbed energy (Supplementary Information; Section SI.7). In the complete cells the dipole bars act as broad, strongly damped antennas that pass energy to the carbon through the near field within about one optical cycle; the near field then carries about two-thirds of the carbon absorption, while most of the additional absorption that the bars introduce between 0.54 and 0.79~$\mu$m is dissipated in their own gold. The spacer increases the energy held transiently by the particles and delays its delivery to the carbon, and it red-shifts the bar resonance from 0.67 to 1.05~$\mu$m, where it raises the carbon share of absorption. All four responses decay within about 10~fs, so the time-integrated partition is fixed by the spectra discussed above.

\subsection{Spacer thickness as a spectral control parameter}

\begin{figure}[!htbp]
    \centering
    \includegraphics[width=1\linewidth]{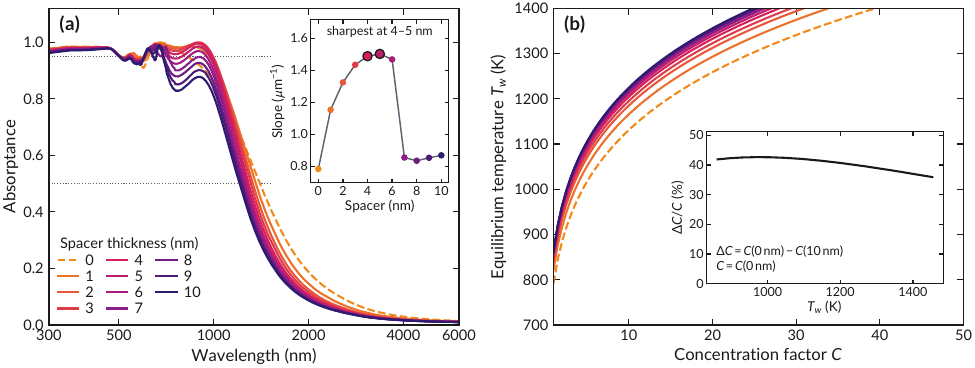}
    \caption{(a) Absorptance spectra of the complete metasurface for spacer thicknesses 0-10~nm; the inset gives the sharpness of the transition against spacer thickness, the mean slope $0.45/(\lambda_{50}-\lambda_{95})$ of the absorptance between the 95\% edge, $\lambda_{95}$, and the half-absorption wavelength, $\lambda_{50}$. (b) Corresponding radiative-limit equilibrium temperature $T_w$ of the fixed Au--carbon optical model as a function of concentration factor $C$ (values near 1350~K are not feasible Au operating temperatures); the inset gives the relative reduction $\Delta C/C$ to reach certain $T_w$, showing nearly 40\% less required concentration at high temperatures. } 
    \label{fig:Fig8}
\end{figure}

Once the coupling mechanism is established, spacer thickness provides a continuous means of tuning how the resonant field is shared between the dielectric gap and the carbon absorber. The sweep in Fig.~\ref{fig:Fig8} exposes the resulting trade-off between solar-band capture and thermal-band rejection.

Across the 1--10~nm range, the solar plateau and the cutoff sharpness evolve in opposite directions. Thicker spacers reduce part of the visible-band response but produce a markedly steeper infrared roll-off with the largest mean slope of 1.50~$\mu$m$^{-1}$ at 4--5~nm compared to 0.78~$\mu$m$^{-1}$ [inset of Fig.~\ref{fig:Fig8}(a)]. 

Inserting these spectra into Eq.~\ref{eq:radiative-limit-balance} converts the optical trade-off into a thermal one [Fig.~\ref{fig:Fig8}(b)]. In the fixed-material radiative model, the calculated temperatures exceed those of the 95\%--10\% reference and the 95\%--5\% profile with spacers above 6~nm, while the thinner spacers approach it  for $T_w\gtrsim1300$~K. This portion of the curve crosses the bulk Au melting point and describes spectral potential only, not the operating regime of this device. At low concentration, the very small absorptivity above 3000~nm further suppresses radiative loss. Crucially, $T_w$ rises monotonically with spacer thickness even as the visible absorptance falls, and the inset quantifies the corresponding relative reduction $\Delta C/C$ in the concentration required to reach a prescribed temperature, 36--43\% between 870 and 1450~K. The result separates two optimisation objectives that are often conflated: maximising integrated solar absorptance and maximising high-temperature radiative performance. A two-level reference profile is useful for locating the desired cutoff, but only the full thermally weighted spectrum captures the effect of a finite transition width and a non-uniform infrared tail.

\subsection{Angular robustness of the spacer-engineered spectrum}

We simulated both polarisations over $0^\circ\leq\theta\leq88^\circ$ and $0.3$--$9~\mu$m. The fourfold in-plane symmetry of the unit cell makes the normal-incidence absorptance independent of the orientation of the incident linear polarisation. The angle-resolved maps and polarisation detail are presented in Figs.~S3 and S9. A separate azimuthal comparison for the 4~nm complete cell is shown in Fig.~S4; the small solar-band differences support using the $\phi=0^\circ$ slice for an approximate hemispherical estimate, while leaving an uncertainty for azimuths and infrared wavelengths not sampled.

We compare two radiative estimates. The normal-incidence reference uses $\theta=0$ absorptance for both the solar and thermal terms. Assuming the residual azimuthal dependence to be negligible for this estimate, we approximate the hemispherical emissivity through
\begin{equation}
    \varepsilon_\mathrm{hemi}(\lambda)
    = 2\int_0^{\pi/2}
      \bar{\alpha}(\lambda,\theta)\cos\theta\sin\theta\,d\theta,
    \label{eq:epsilon-hemi}
\end{equation}
where $\bar{\alpha}=\tfrac{1}{2}(\alpha_P+\alpha_S)$. Here, Kirchhoff's law in Eq.~\ref{eq:kirchhoff-law} equates absorptance from an incoming incident plane wave with $\mathbf{n}(\theta,\phi)$, $\phi$ being the azimuth of its in-plane wavevector, with the emissivity into the time-reversed outgoing channel $\mathbf{n}(\theta,\phi+180^\circ)$ at the same wavelength and polarisation. The $C_4$ symmetry of the unit cell makes the two azimuth labels equivalent, so no direction reversal enters Eq.~\ref{eq:epsilon-hemi}. The sampled azimuthal dependence is small in the solar-band average (Fig.~S4); $\bar{\alpha}$ is evaluated at $\phi=0^\circ$ throughout. This does not constitute a complete azimuthal error bound for thermal emission. Over the concentration range considered here, the concentrated solar cone remains narrow compared with the angular range relevant to hemispherical thermal emission. We therefore approximate the solar absorptance by its normal-incidence value, $\alpha_a(\lambda)=\alpha(\lambda,0)$, while retaining the simulated angular dependence in the thermal term,
\begin{equation}
    C\int_0^\infty
    \alpha_a(\lambda)I_{\lambda\odot}(\lambda)\,d\lambda
    =
    \pi\int_0^\infty
    \varepsilon_\mathrm{hemi}(\lambda)
    \left[B_\lambda(\lambda,T_w)-B_\lambda(\lambda,T_0)\right]d\lambda.
    \label{eq:balance-lambertian}
\end{equation}
The hemispherical emissivity is evaluated using the simulated polar sweep to $88^\circ$ and spectra extending to $9~\mu$m. The finite spectral range and sampled azimuth are limitations of the numerical estimate, particularly at lower temperatures. Under the $\cos\theta\sin\theta$ projection, the polarisation average entering $\varepsilon_\mathrm{hemi}$ varies smoothly with angle, and the hemispherical estimate lies 32--91~K below the normal-incidence one at $C=1$, and between 58~K below and 5~K above it at $C=50$. 

\begin{figure}[!htbp]
    \centering
    \includegraphics[width=\linewidth]{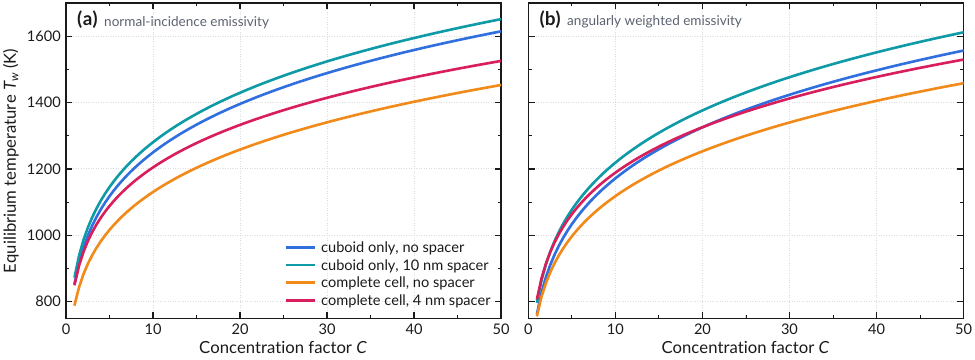}
    \caption{Radiative-limit equilibrium temperature $T_w$ versus concentration factor $C$ for the four metasurface designs. (a) Normal-incidence absorptance used for both solar absorption and thermal emission. (b) Approximate angularly weighted emissivity from Eq.~\ref{eq:epsilon-hemi} used for thermal emission, with normal-incidence absorptance retained for the solar input. Values above the bulk Au melting point are formal radiative-limit benchmarks, not Au operating temperatures. Absorptance beyond 9~$\mu$m is held at its 9~$\mu$m value.} 
    \label{fig:ch6:T-vs-C-angular}
\end{figure}

Angular weighting lowers the absolute temperatures, most strongly for the cuboid-only cells at low concentration (the complete cells gain 4--5~K at $C=50$), and preserves the normal-incidence ordering of the four designs above $C\approx20$; below it the complete cell with a 4~nm spacer overtakes the cuboid-only cell without spacer, and below $C\approx2.4$ also the one with a 10~nm spacer [Fig.~\ref{fig:ch6:T-vs-C-angular}]. The cuboid-only structures reach the highest radiative-limit temperatures because their shorter cutoff suppresses thermal emission more strongly, whereas the complete unit cells absorb more solar power but radiate over a broader band. The distinction is physical rather than contradictory: equilibrium temperature measures radiative retention, not total absorbed or extractable heat. The preferred geometry therefore depends on whether the objective is to maximise stagnation temperature, absorbed solar flux or useful thermal output. Within the present angular approximation, the spacer-mediated spectral response remains a robust feature rather than a narrow normal-incidence effect.

\subsection{Phenomenological spectral figure of merit}

\begin{figure}[!htbp]
    \centering
    \includegraphics[width=1\linewidth]{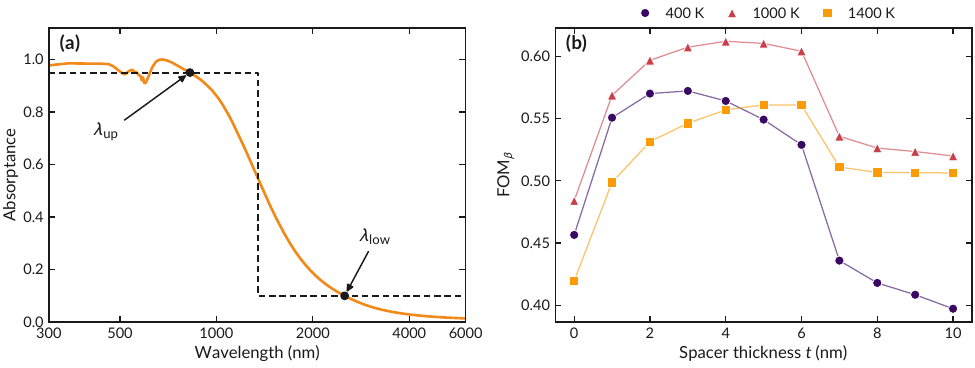}
    \caption{(a) Definition of the threshold wavelengths    $\lambda_{\mathrm{up}}$ and $\lambda_{\mathrm{low}}$ relative to the reference levels $\alpha_{\mathrm{up}}$ and $\alpha_{\mathrm{low}}$. (b) Phenomenological $\mathrm{FOM}_{\beta}$ as a function of spacer thickness at $T_w=400$, 1000 and 1400~K, using the prescribed     temperature-dependent calibration of $\beta$ with $T_{\mathrm{ref}}=830$~K.}
    \label{fig:FOMbeta}
\end{figure}

For rapid comparison across the spacer-thickness series, we introduce a bounded phenomenological figure of merit, $\mathrm{FOM}_{\beta}$, that combines solar-band absorption, infrared suppression and transition sharpness. Equilibrium temperatures and design rankings are determined by the full radiative balance in Eq.~\ref{eq:radiative-limit-balance}; the metric summarises trends within the present design family. Its first component is the solar-weighted absorptivity

\begin{equation}
\overline{\alpha}_{\mathrm{s}} =
\frac{
\displaystyle \int_{\lambda_{\min}}^{\lambda_c}
\alpha(\lambda) I_{\lambda\odot}(\lambda)\,\mathrm{d}\lambda
}{
\displaystyle \int_{\lambda_{\min}}^{\lambda_c}
I_{\lambda\odot}(\lambda)\,\mathrm{d}\lambda
},
\end{equation}
where \(I_{\lambda\odot}(\lambda)\) is the AM1.5D spectrum and \(\lambda_c\) is the reference cutoff. The complementary thermal term is the Planck-weighted absorptivity above the cutoff,

\begin{equation}
\overline{\alpha}_{\mathrm{IR}} =
\frac{
\displaystyle \int_{\lambda_c}^{\lambda_{\max}}
\alpha(\lambda) B_\lambda(\lambda,T_w)\,\mathrm{d}\lambda
}{
\displaystyle \int_{\lambda_c}^{\lambda_{\max}}
B_\lambda(\lambda,T_w)\,\mathrm{d}\lambda
},
\label{eq:alpha_IR}
\end{equation}
where \(B_\lambda(\lambda,T)\) is the blackbody spectral radiance at the evaluation temperature \(T_w\), and \(\lambda_{\max}\) is the upper wavelength limit considered in the calculation, $\lambda_{\max}=6\,\mu\mathrm{m}$. For Fig.~\ref{fig:FOMbeta}, we use $\lambda_{\min}=300$~nm, $\lambda_c=1350$~nm, $\lambda_{\max}=6~\mu$m, $\alpha_{\mathrm{up}}=0.95$ and $\alpha_{\mathrm{low}}=0.10$, matching the two-level reference profile. The normalised integrals ensure that, for any passive absorber with $0\leq\alpha(\lambda)\leq1$, both $\overline{\alpha}_{\mathrm{s}}$ and $\overline{\alpha}_{\mathrm{IR}}$ remain dimensionless and bounded between zero and unity for any finite interval $[\lambda_{\min},\lambda_{\max}]$. At lower operating temperatures the Planck spectrum shifts towards longer wavelengths. In the present design family, however, all simulated spectra have already approached a small, slowly varying and nearly common asymptotic absorptance by approximately $6~\mu\mathrm{m}$. The omitted long-wavelength tail is therefore expected to contribute similarly across the spacer-thickness series and to provide little additional discrimination between these designs. This point is particularly relevant at $T_w=400~\mathrm{K}$, where the thermal spectrum lies predominantly beyond the solar-to-infrared transition: in this regime $\overline{\alpha}_{\mathrm{IR}}$ is only weakly discriminating, and the figure of merit is governed primarily by the solar-band term and the position of the spectral transition. As $T_w$ increases, the Planck weighting moves towards shorter wavelengths and the infrared term becomes progressively more sensitive to differences among the spectra.

The factor $S_{\beta}$ characterises the position and width of the main transition from the high-absorptivity solar band to the low-emissivity infrared band. Two threshold wavelengths are extracted from the absorptance spectrum on the two sides of the target cutoff $\lambda_c$, as illustrated in Fig.~\ref{fig:FOMbeta}(a). On the short-wavelength side of $\lambda_c$, $\lambda_{\mathrm{up}}$ is defined as the last wavelength at which the spectrum crosses the high-absorptivity level $\alpha_{\mathrm{up}}$. On the long-wavelength side, $\lambda_{\mathrm{low}}$ is defined as the first wavelength at which the spectrum crosses the low-absorptivity level $\alpha_{\mathrm{low}}$. The interval $\lambda_{\mathrm{up}}\leq\lambda\leq\lambda_{\mathrm{low}}$ therefore measures the spectral width over which the absorber changes from the prescribed high-absorption state to the prescribed low-emission state.  Selecting the last and first crossings on the two sides of $\lambda_c$ provides a consistent definition of the transition width, although the resulting thresholds may remain sensitive to individual resonances near the prescribed absorptivity levels.

The relative emphasis placed on solar-band absorption and infrared suppression is controlled by a positive dimensionless parameter $\beta$. We therefore define a one-parameter family of spectral-selectivity factors,
\begin{equation}
    S_{\beta}
    =
    \frac{1}{2}
    \left[
        \left(
            \frac{\lambda_{\mathrm{up}}}{\lambda_c}
        \right)^{\beta}
        +
        \left(
            \frac{\lambda_c}{\lambda_{\mathrm{low}}}
        \right)^{1/\beta}
    \right],
    \qquad \beta>0.
\end{equation}
For $\lambda_{\mathrm{up}}\leq\lambda_c\leq\lambda_{\mathrm{low}}$, both ratios lie in $[0,1]$, and therefore $0\leq S_{\beta}\leq1$. The first term measures how far strong absorption persists towards the target cutoff from the solar side, while the second measures how rapidly the spectrum reaches the required low-emissivity level beyond the cutoff. The neutral choice $\beta=1$ weights the solar and infrared sides symmetrically. Values $\beta>1$ place greater emphasis on sustaining high absorptivity towards the cutoff, whereas $0<\beta<1$ place greater emphasis on reaching the low-emissivity regime rapidly beyond the cutoff.
The corresponding one-parameter family of figures of merit is
\begin{equation}
    \mathrm{FOM}_{\beta}
    =
    \overline{\alpha}_{\mathrm{s}}^{\,\beta}
    \left(
        1-\overline{\alpha}_{\mathrm{IR}}
    \right)^{1/\beta}
    S_{\beta}.
    \label{eq:FOMbeta}
\end{equation}
For the calculations presented below, we use the temperature-dependent prescription
\begin{equation}
    \beta(T_w)
    =
    \frac{T_{\mathrm{ref}}}{T_w},
\end{equation}
where $T_{\mathrm{ref}}=830~\mathrm{K}$ is the reference temperature at which $\beta=1$. The reference-cutoff map in Fig.~\ref{fig:target-absorber}(b) shows an approximate crossover near this temperature: below it the equilibrium temperature depends only weakly on the cutoff wavelength $\lambda_c$ over $1000$--$2000$~nm, whereas at higher temperatures the cutoff becomes more influential. This prescription progressively shifts the weighting from solar capture at lower temperatures towards infrared suppression at higher temperatures. It is a phenomenological calibration motivated by the map, not a result derived from the radiative-balance equation. The same prescription is applied to every spectrum compared at a given evaluation temperature.

The multiplicative form of $\mathrm{FOM}_{\beta}$ ensures that poor performance in any one of the three essential aspects, solar-band absorption, infrared-emission suppression, or transition sharpness and position, cannot be fully compensated by another. Since each factor is bounded between zero and unity, the metric also satisfies $0\leq\mathrm{FOM}_{\beta}\leq1$. For the ideal step absorber with unit absorptivity below $\lambda_c$, zero absorptivity above $\lambda_c$, and an abrupt transition such that $\lambda_{\mathrm{up}}=\lambda_c=\lambda_{\mathrm{low}}$, one obtains $\overline{\alpha}_{\mathrm{s}}=1$, $\overline{\alpha}_{\mathrm{IR}}=0$, $S_{\beta}=1$, and therefore $\mathrm{FOM}_{\beta}=1$ for every $\beta>0$. The metric thus recovers the desired ideal limit while retaining sensitivity to the finite plateau values and transition width of realistic spectra.

Applied to the spacer sweep, $\mathrm{FOM}_{\beta}$ is evaluated for spacer thicknesses from 0 to 10~nm at target temperatures of 400, 1000 and 1400~K [Fig.~\ref{fig:FOMbeta}(b)]. At 400~K the metric is largest at 3~nm, closely following the saturation of the absorptance below $\lambda_c$. The decrease at spacer thicknesses of 7–10 nm follows $\lambda_{\mathrm{up}}$, which falls to about 700~nm once the resonance near 900~nm no longer reaches $\alpha_{\mathrm{up}}$ (Fig.~\ref{fig:Fig8}). This low-temperature behaviour is consistent with the reference-cutoff map in Fig.~\ref{fig:target-absorber}(b), where the radiative-limit equilibrium is only weakly sensitive to the precise cutoff position below $T_{\mathrm{ref}}$. For the present design family, the spectra have also converged to a similarly low absorptance in the long-wavelength infrared, so the infrared term provides little additional discrimination at 400~K and the FOM is governed primarily by solar-band absorption.

At 1000 and 1400~K, differences in the spectral transition and infrared response become more important, and the optimum lies at 4~nm (1000~K) and 5~nm (1400~K), where the cutoff is sharper and the infrared emissivity is lower. The 1400~K curve is a formal comparison of spectra above the bulk Au melting point; it does not represent an operating temperature for this structure. These trends are consistent with the spacer-thickness sweep in Fig.~\ref{fig:Fig8} and illustrate the shift from solar-absorption-dominated ranking at low temperature towards greater sensitivity to infrared suppression and transition sharpness. Within the present design family, $\mathrm{FOM}_{\beta}$ provides a compact view of this trade-off. Its calibration and ranking would need separate validation for other absorber architectures, and the metric does not establish useful-heat efficiency or material stability.

\section{Conclusion}
Separating a dielectric gap from the absorbing carbon layer gives a physical means of controlling loss in a metal--carbon selective metasurface. The spacer redistributes the localised field and carbon absorption, broadens the useful solar band, and changes the infrared transition. A controlled sequence from a planar stack to the complete unit cell, together with material-resolved loss, supports this mechanism.

The radiative calculation shows why solar-weighted absorption alone does not rank spectra at elevated temperature: transition sharpness and residual infrared emissivity can dominate the radiative-limit equilibrium. Approximate hemispherical weighting lowers the equilibrium temperatures at low concentration and preserves the ordering of the compared designs at high concentration. A phenomenological figure of merit captures the temperature-dependent spacer trend within this design family, while the full radiative balance determines the equilibrium temperatures. The high-temperature values are benchmarks obtained from a fixed, temperature-independent optical model under idealised radiative conditions. Practical operation, and any claim of useful solar-thermal output, require a thermally durable implementation, temperature-dependent optical data and a balance including heat extraction and parasitic losses.

\paragraph{Supplementary Information.}
The Supplementary Information is included at the end of this document, following the references. It provides the optical constants and material-resolved absorptance, angular and polarisation-resolved spectra, ideal-step and field-profile analyses, time-domain energy flow, and numerical convergence tests (Supplementary Sections SI.1--SI.9; Figs. S1--S10).

\printbibliography

@article{lambertz2026,
  author       = {Lambertz, A. and Lee, N. and Vafa, A. P. and Shautsova, V. and Xu, F. and Song, J.-H. and G{\"u}sken, N. A. and Selvin, S. and Florescu, M. and Brongersma, M. L. and van de Groep, J. and Alarc{\'o}n-Llad{\'o}, E.},
  title        = {Integrating hyperuniform disordered light trapping layers in ultrathin {c-Si} solar cells},
  journal      = {ACS Photonics},
  volume       = {13},
  number       = {20},
  pages        = {6104--6113},
  year         = {2026},
  doi          = {10.1021/acsphotonics.6c01224}
}

@article{azadMetasurfaceBroadbandSolar2016,
  author       = {Azad, A. K. and Kort-Kamp, W. J. M. and Sykora, M. and Weisse-Bernstein, N. R. and Luk, T. S. and Taylor, A. J. and Dalvit, D. A. R. and Chen, H.-T.},
  title        = {Metasurface broadband solar absorber},
  journal      = {Sci. Rep.},
  volume       = {6},
  pages        = {20347},
  year         = {2016},
  doi          = {10.1038/srep20347}
}

@misc{nistGoldPoint,
  author       = {{National Institute of Standards and Technology}},
  title        = {{Kelvin}: {ITS-90}},
  howpublished = {\url{https://www.nist.gov/si-redefinition/kelvin/kelvin-its-90}},
  note         = {Gold freezing point: 1337.33 K; accessed 2026-09-28}
}

@article{abed2017,
  author       = {Abedini Dereshgi, S. and Ghobadi, A. and Hajian, H. and Butun, B. and Ozbay, E.},
  title        = {Ultra-broadband, lithography-free, and large-scale compatible perfect absorbers: the optimum choice of metal layers in metal-insulator multilayer stacks},
  journal      = {Sci. Rep.},
  volume       = {7},
  number       = {1},
  pages        = {14872},
  year         = {2017},
  doi          = {10.1038/s41598-017-13837-8}
}

@article{ahmadUltrathinMetalFilms2012,
  author       = {Ahmad, N. and Stokes, J. and Fox, N. A. and Teng, M. and Cryan, M. J.},
  title        = {Ultra-thin metal films for enhanced solar absorption},
  journal      = {Nano Energy},
  volume       = {1},
  number       = {6},
  pages        = {777--782},
  year         = {2012},
  doi          = {10.1016/j.nanoen.2012.08.004}
}

@article{ghobadi97PercentLight2017,
  author       = {Ghobadi, A. and Dereshgi, S. A. and Hajian, H. and Birant, G. and Butun, B. and Bek, A. and Ozbay, E.},
  title        = {97 Percent light absorption in an ultrabroadband frequency range utilizing an ultrathin metal layer: randomly oriented, densely packed dielectric nanowires as an excellent light trapping scaffold},
  journal      = {Nanoscale},
  volume       = {9},
  number       = {43},
  pages        = {16652--16660},
  year         = {2017},
  doi          = {10.1039/C7NR04186A}
}

@article{hanBroadbandSolarThermal2016,
  author       = {Han, S. and Shin, J.-H. and Jung, P.-H. and Lee, H. and Lee, B. J.},
  title        = {Broadband solar thermal absorber based on optical metamaterials for high-temperature applications},
  journal      = {Adv. Opt. Mater.},
  volume       = {4},
  number       = {8},
  pages        = {1265--1273},
  year         = {2016},
  doi          = {10.1002/adom.201600236}
}

@article{kalogirouSolarThermalCollectors2004,
  author       = {Kalogirou, S. A.},
  title        = {Solar thermal collectors and applications},
  journal      = {Prog. Energy Combust. Sci.},
  volume       = {30},
  number       = {3},
  pages        = {231--295},
  year         = {2004},
  doi          = {10.1016/j.pecs.2004.02.001}
}

@article{lalUsingSpacerLayers2012,
  author       = {Lal, N. N. and Zhou, H. and Hawkeye, M. and Sinha, J. K. and Bartlett, P. N. and Amaratunga, G. A. J. and Baumberg, J. J.},
  title        = {Using spacer layers to control metal and semiconductor absorption in ultrathin solar cells with plasmonic substrates},
  journal      = {Phys. Rev. B},
  volume       = {85},
  number       = {24},
  pages        = {245318},
  year         = {2012},
  doi          = {10.1103/PhysRevB.85.245318}
}

@article{linStructuredGrapheneMetamaterial2020,
  author       = {Lin, K.-T. and Lin, H. and Yang, T. and Jia, B.},
  title        = {Structured graphene metamaterial selective absorbers for high efficiency and omnidirectional solar thermal energy conversion},
  journal      = {Nat. Commun.},
  volume       = {11},
  number       = {1},
  pages        = {1389},
  year         = {2020},
  doi          = {10.1038/s41467-020-15116-z}
}

@article{pillaiEffectDielectricSpacer2011,
  author       = {Pillai, S. and Beck, F. J. and Catchpole, K. R. and Ouyang, Z. and Green, M. A.},
  title        = {The effect of dielectric spacer thickness on surface plasmon enhanced solar cells for front and rear side depositions},
  journal      = {J. Appl. Phys.},
  volume       = {109},
  number       = {7},
  pages        = {073105},
  year         = {2011},
  doi          = {10.1063/1.3567299}
}

@article{rheeMetamaterialbasedPerfectAbsorbers2014,
  author       = {Rhee, J. Y. and Yoo, Y. J. and Kim, K. W. and Kim, Y. J. and Lee, Y. P.},
  title        = {Metamaterial-based perfect absorbers},
  journal      = {J. Electromagn. Waves Appl.},
  volume       = {28},
  number       = {13},
  pages        = {1541--1580},
  year         = {2014},
  doi          = {10.1080/09205071.2014.944273}
}

@article{tianReviewSolarCollectors2013,
  author       = {Tian, Y. and Zhao, C. Y.},
  title        = {A review of solar collectors and thermal energy storage in solar thermal applications},
  journal      = {Appl. Energy},
  volume       = {104},
  pages        = {538--553},
  year         = {2013},
  doi          = {10.1016/j.apenergy.2012.11.051}
}

@article{tinchevThermalSolarAbsorber2010,
  author       = {Tinchev, S. S. and Nikolova, P. I. and Dyulgerska, Y. T.},
  title        = {Thermal solar absorber made of diamond-like carbon thin films},
  journal      = {J. Phys.: Conf. Ser.},
  volume       = {223},
  number       = {1},
  pages        = {012017},
  year         = {2010},
  doi          = {10.1088/1742-6596/223/1/012017}
}

@article{tomescuNanopillarsMetasurfaceModelled2020,
  author       = {Tomescu, R. and Kusko, C. and Cristea, D. and Calinoiu, R. and Parvulescu, C.},
  title        = {Nano-pillars metasurface modelled for perfect absorption at specific wavelengths in infrared spectral regime},
  journal      = {Solid State Electron. Lett.},
  volume       = {2},
  pages        = {146--150},
  year         = {2020},
  doi          = {10.1016/j.ssel.2020.11.002}
}

@article{wanBroadbandMetasurfaceAbsorber2015,
  author       = {Wan, C. and Chen, L. and Cryan, M. J.},
  title        = {Broadband metasurface absorber for solar thermal applications},
  journal      = {J. Opt.},
  volume       = {17},
  number       = {12},
  pages        = {125103},
  year         = {2015},
  doi          = {10.1088/2040-8978/17/12/125103}
}

@article{wangHighlyEfficientSelective2015,
  author       = {Wang, H. and Prasad Sivan, V. and Mitchell, A. and Rosengarten, G. and Phelan, P. and Wang, L.},
  title        = {Highly efficient selective metamaterial absorber for high-temperature solar thermal energy harvesting},
  journal      = {Sol. Energy Mater. Sol. Cells},
  volume       = {137},
  pages        = {235--242},
  year         = {2015},
  doi          = {10.1016/j.solmat.2015.02.019}
}

@article{wangPerfectSelectiveMetamaterial2013a,
  author       = {Wang, H. and Wang, L.},
  title        = {Perfect selective metamaterial solar absorbers},
  journal      = {Opt. Express},
  volume       = {21},
  number       = {S6},
  pages        = {A1078--A1093},
  year         = {2013},
  doi          = {10.1364/OE.21.0A1078}
}

@article{wanSelectiveMetasurfaceAbsorber2016,
  author       = {Wan, C. and Ho, Y. and Nunez-Sanchez, S. and Chen, L. and Lopez-Garcia, M. and Pugh, J. and Zhu, B. and Selvaraj, P. and Mallick, T. and Senthilarasu, S. and Cryan, M. J.},
  title        = {A selective metasurface absorber with an amorphous carbon interlayer for solar thermal applications},
  journal      = {Nano Energy},
  volume       = {26},
  pages        = {392--397},
  year         = {2016},
  doi          = {10.1016/j.nanoen.2016.05.013}
}

@article{xuNearperfectSpectrallyselectiveMetasurface2022,
  author       = {Xu, M. and Guo, L. and Zhang, P. and Qiu, Y. and Li, Q. and Wang, J.},
  title        = {Near-perfect spectrally-selective metasurface solar absorber based on tungsten octagonal prism array},
  journal      = {RSC Adv.},
  volume       = {12},
  number       = {26},
  pages        = {16823--16834},
  year         = {2022},
  doi          = {10.1039/D2RA02802F}
}

@article{zhangSolarSelectiveAbsorber2022,
  author       = {Zhang, J. and Wang, C. and Shi, J. and Wei, D. and Zhao, H. and Ma, C.},
  title        = {Solar selective absorber for emerging sustainable applications},
  journal      = {Adv. Energy Sustainability Res.},
  volume       = {3},
  number       = {3},
  pages        = {2100195},
  year         = {2022},
  doi          = {10.1002/aesr.202100195}
}

@article{Schwede2010Photon-enhancedSystems,
  author       = {Schwede, J. W. and Bargatin, I. and Riley, D. C. and Hardin, B. E. and Rosenthal, S. J. and Sun, Y. and Schmitt, F. and Pianetta, P. and Howe, R. T. and Shen, Z. X. and Melosh, N. A.},
  title        = {Photon-enhanced thermionic emission for solar concentrator systems},
  journal      = {Nat. Mater.},
  volume       = {9},
  number       = {9},
  pages        = {762--767},
  year         = {2010},
  doi          = {10.1038/nmat2814}
}

@article{Mao2020ManipulatingReflection,
  author       = {Mao, P. and Liu, C. and Song, F. and Han, M. and Maier, S. A. and Zhang, S.},
  title        = {Manipulating disordered plasmonic systems by external cavity with transition from broadband absorption to reconfigurable reflection},
  journal      = {Nat. Commun.},
  volume       = {11},
  number       = {1},
  pages        = {1538},
  year         = {2020},
  doi          = {10.1038/s41467-020-15349-y}
}

@article{baffouThermoplasmonicsMetallicNanostructures2013,
  author       = {Baffou, G. and Quidant, R.},
  title        = {Thermo-plasmonics of metal nanostructures},
  journal      = {Laser Photonics Rev.},
  volume       = {7},
  number       = {2},
  pages        = {171--187},
  year         = {2013},
  doi          = {10.1002/lpor.201200019}
}

@book{incroperaFundamentalsHeatMass2017,
  author       = {Bergman, T. L. and Lavine, A. S. and Incropera, F. P. and DeWitt, D. P.},
  title        = {Fundamentals of heat and mass transfer},
  publisher    = {John Wiley \& Sons},
  edition      = {7th},
  isbn         = {978-0470501979},
  year         = {2011}
}

@book{modestRadiativeHeatTransfer2021,
  author       = {Modest, M. F. and Mazumder, S.},
  title        = {Radiative heat transfer},
  publisher    = {Academic Press},
  edition      = {4th},
  isbn         = {978-0128181546},
  year         = {2021}
}

@article{fanThermalPhotonicsEnergy2017,
  author       = {Fan, S.},
  title        = {Thermal photonics and energy applications},
  journal      = {Joule},
  volume       = {1},
  number       = {2},
  pages        = {264--273},
  year         = {2017},
  doi          = {10.1016/j.joule.2017.09.012}
}

@article{liNanophotonicControlThermal2018,
  author       = {Li, W. and Fan, S.},
  title        = {Nanophotonic control of thermal radiation for energy applications},
  journal      = {Opt. Express},
  volume       = {26},
  number       = {12},
  pages        = {15995--16021},
  year         = {2018},
  doi          = {10.1364/OE.26.015995}
}

@article{vafaBalancingOrderDisorder2026,
  author       = {Vafa, A. P. and Silva, S. R. P. and Florescu, M.},
  title        = {Balancing order and disorder: optimal light trapping in ultrathin silicon photovoltaics},
  journal      = {Optica},
  volume       = {13},
  number       = {6},
  pages        = {1118--1129},
  year         = {2026},
  doi          = {10.1364/OPTICA.592268}
}

@article{tavakoliOver65Sunlight2022,
  author       = {Tavakoli, N. and Spalding, R. and Lambertz, A. and Koppejan, P. and Gkantzounis, G. and Wan, C. and R{\"o}hrich, R. and Kontoleta, E. and Koenderink, A. F. and Sapienza, R. and Florescu, M. and Alarc{\'o}n-Llad{\'o}, E.},
  title        = {Over 65\% sunlight absorption in a 1~$\mu$m {Si} slab with hyperuniform texture},
  journal      = {ACS Photonics},
  volume       = {9},
  number       = {4},
  pages        = {1206--1217},
  year         = {2022},
  doi          = {10.1021/acsphotonics.1c01668}
}

@article{schulerThermalEmissionFinite2009,
  author       = {Schuler, C. J. and Wolff, C. and Busch, K. and Florescu, M.},
  title        = {Thermal emission from finite photonic crystals},
  journal      = {Appl. Phys. Lett.},
  volume       = {95},
  number       = {24},
  pages        = {241103},
  year         = {2009},
  doi          = {10.1063/1.3275578}
}

@article{florescuImprovingSolarCell2007,
  author       = {Florescu, M. and Lee, H. and Puscasu, I. and Pralle, M. and Florescu, L. and Ting, D. Z. and Dowling, J. P.},
  title        = {Improving solar cell efficiency using photonic band-gap materials},
  journal      = {Sol. Energy Mater. Sol. Cells},
  volume       = {91},
  number       = {17},
  pages        = {1599--1610},
  year         = {2007},
  doi          = {10.1016/j.solmat.2007.05.001}
}

@article{Steiner2016RatingRobins,
  author       = {Steiner, M. and Siefer, G. and Baudrit, M. and Askins, S. and Dom{\'i}nguez, C. and Ant{\'o}n, I. and Roca, F. and Fucci, R. and Cancro, C. and Romano, A. and Graditi, G. and Pugliatti, P. M. and Di Stefano, A. and Kenny, R. and Norton, M. and Minuto, A. and Morabito, P. and Muller, M. and Riley, D. and Pratt, L.},
  title        = {Rating of {CPV} modules: results of module round robins},
  journal      = {AIP Conf. Proc.},
  volume       = {1766},
  number       = {1},
  pages        = {040005},
  year         = {2016},
  doi          = {10.1063/1.4962082}
}

@article{babarWeaver2015,
 author  = {Olmon, Robert L. and Slovick, Brian and
             Johnson, Timothy W. and Shelton, David and
             Oh, Sang-Hyun and Boreman, Glenn D. and
             Raschke, Markus B.},
  title   = {Optical dielectric function of gold},
  journal = {Physical Review B},
  volume  = {86},
  number  = {23},
  pages   = {235147},
  year    = {2012},
  doi     = {10.1103/PhysRevB.86.235147},
  url     = {https://doi.org/10.1103/PhysRevB.86.235147}
}

@article{hagemannOpticalConstantsFar1975,
   author       = {Babar, S. and Weaver, J. H.},
  title        = {Optical constants of {Cu}, {Ag}, and {Au} revisited},
  journal      = {Appl. Opt.},
  volume       = {54},
  number       = {3},
  pages        = {477--481},
  year         = {2015},
  doi          = {10.1364/AO.54.000477}
}

\clearpage
\setcounter{figure}{0}
\renewcommand{\thefigure}{S\arabic{figure}}
\renewcommand{\theHfigure}{supp.\arabic{figure}}
\setcounter{table}{0}
\renewcommand{\thetable}{S\arabic{table}}
\renewcommand{\theHtable}{supp.\arabic{table}}
\setcounter{equation}{0}
\renewcommand{\theequation}{S\arabic{equation}}
\renewcommand{\theHequation}{supp.\arabic{equation}}
\setcounter{section}{0}
\renewcommand{\thesection}{SI.\arabic{section}}
\renewcommand{\theHsection}{supp.\arabic{section}}
\setcounter{secnumdepth}{1}

\begin{center}
    {\Large\bfseries Supplementary Information}\\[0.65em]
    {\large Spacer-Engineered Spectral Selectivity\\ in Metal--Carbon Solar--Thermal Absorbers}\\[0.75em]
    X. Romain\textsuperscript{1}, A. P. Vafa\textsuperscript{1,2}, and M. Florescu\textsuperscript{1,*}\\[0.5em]
    {\small \textsuperscript{1}Optoelectronics Research Centre, University of Southampton, Southampton, UK}\\
    {\small \textsuperscript{2}Advanced Technology Institute, University of Surrey, Guildford, UK}\\[0.5em]
    {\small *Email: m.florescu@soton.ac.uk}
\end{center}
\vspace{1em}

\section{Optical constants}
\begin{figure}[htbp]
\centering\includegraphics[width=\linewidth]{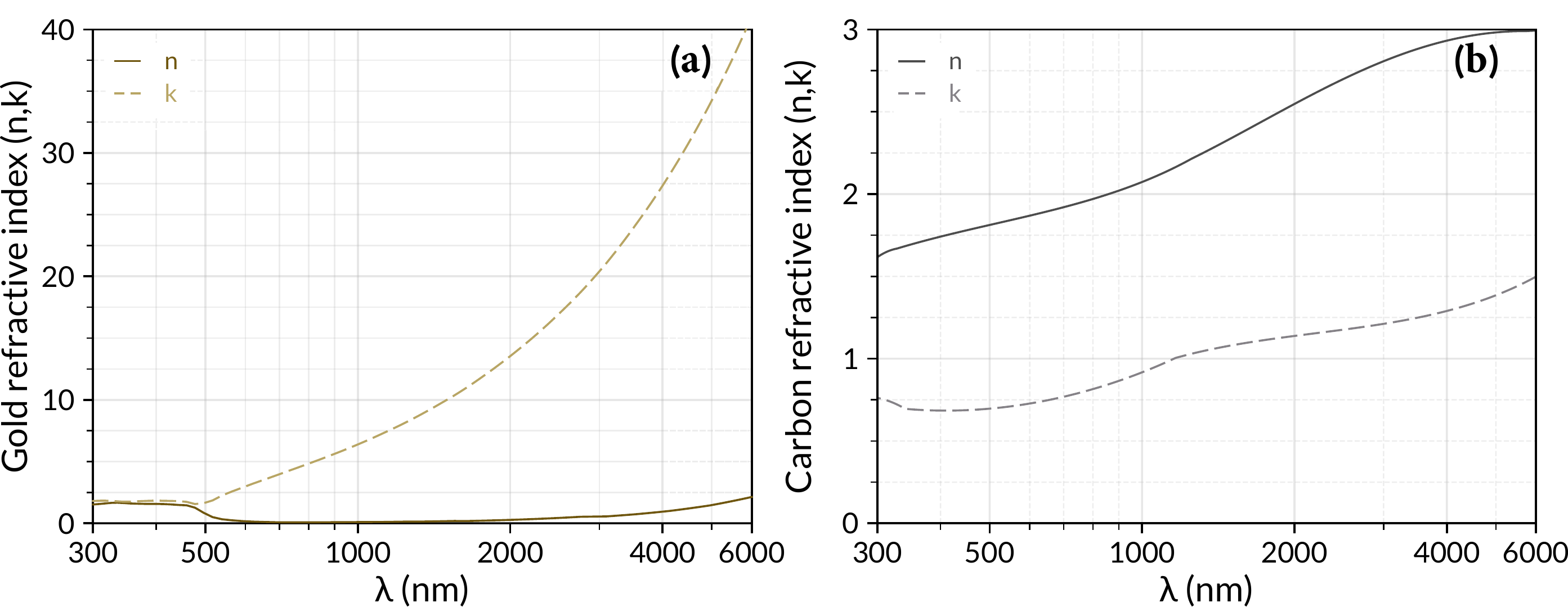}
\caption{Optical constants used in the electromagnetic simulations. (a) Real and imaginary parts of the refractive index of gold. (b) Corresponding refractive-index data for amorphous carbon.}\label{fig:S1}
\end{figure}

The gold and amorphous-carbon optical constants used in the electromagnetic simulations are shown in Fig.~\ref{fig:S1}. The data are taken from Refs.~\cite{babarWeaver2015,hagemannOpticalConstantsFar1975} and are treated as temperature independent. The spacer is modelled as an idealised lossless dielectric with real refractive index 1.45, representative of a SiO$_2$-like optical gap rather than a dispersive model of physical SiO$_2$. Because the model excludes temperature dependence and morphological change, none of the calculated equilibrium temperatures validates thermal durability of the Au device. Bulk gold melts at 1337.33~K~\cite{nistGoldPoint}.

\clearpage

\section{Material-resolved absorptance of the intermediate structures}
\begin{figure}[htbp]
\centering
\begin{subfigure}{0.32\linewidth}\centering\includegraphics[width=\linewidth]{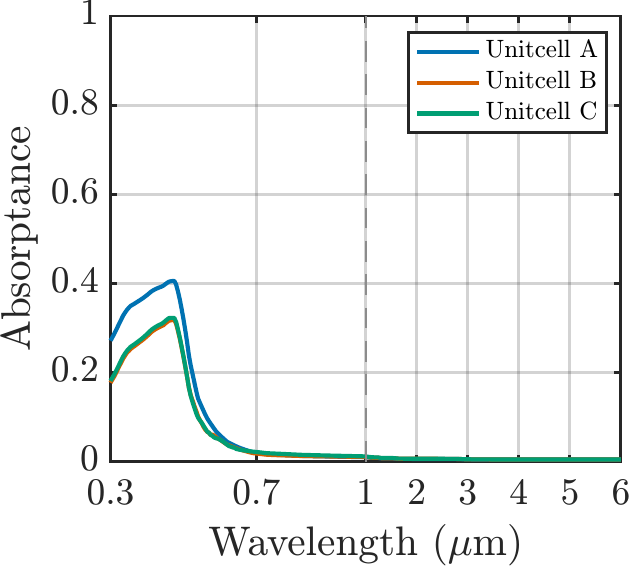}\caption{}\end{subfigure}\hfill
\begin{subfigure}{0.32\linewidth}\centering\includegraphics[width=\linewidth]{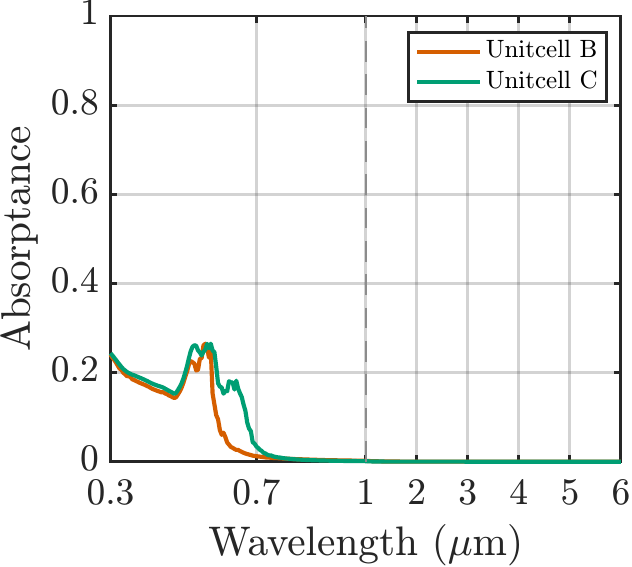}\caption{}\end{subfigure}\hfill
\begin{subfigure}{0.32\linewidth}\centering\includegraphics[width=\linewidth]{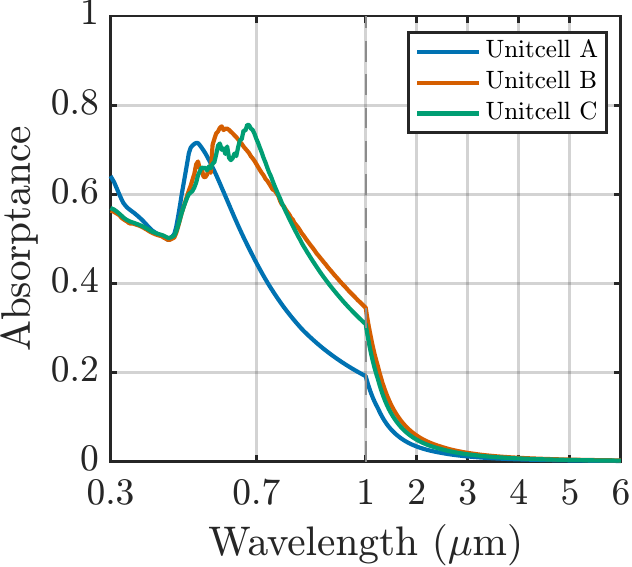}\caption{}\end{subfigure}
\caption{Material-resolved absorptance of the intermediate structures in (a) the gold reflector, (b) the patterned gold resonator and (c) the amorphous-carbon film.}\label{fig:S2}
\end{figure}

The absorbed power density is integrated separately over the gold reflector, patterned gold resonator and amorphous-carbon film (Fig.~\ref{fig:S2}). The cuboid lowers reflector and carbon absorption in the short visible below 500~nm, where gold is very lossy, and raises carbon absorption beyond that (up to 0.13 near 760~nm) by additional coupling. The spacer adds a shoulder to the gold resonator profile at 560~nm and steepens the carbon profile roll-off by controlling the coupling rate between the resonator near-field and carbon.

\clearpage

\section{Polarisation-resolved angular response}
\begin{figure}[htbp]
\centering\includegraphics[width=\linewidth]{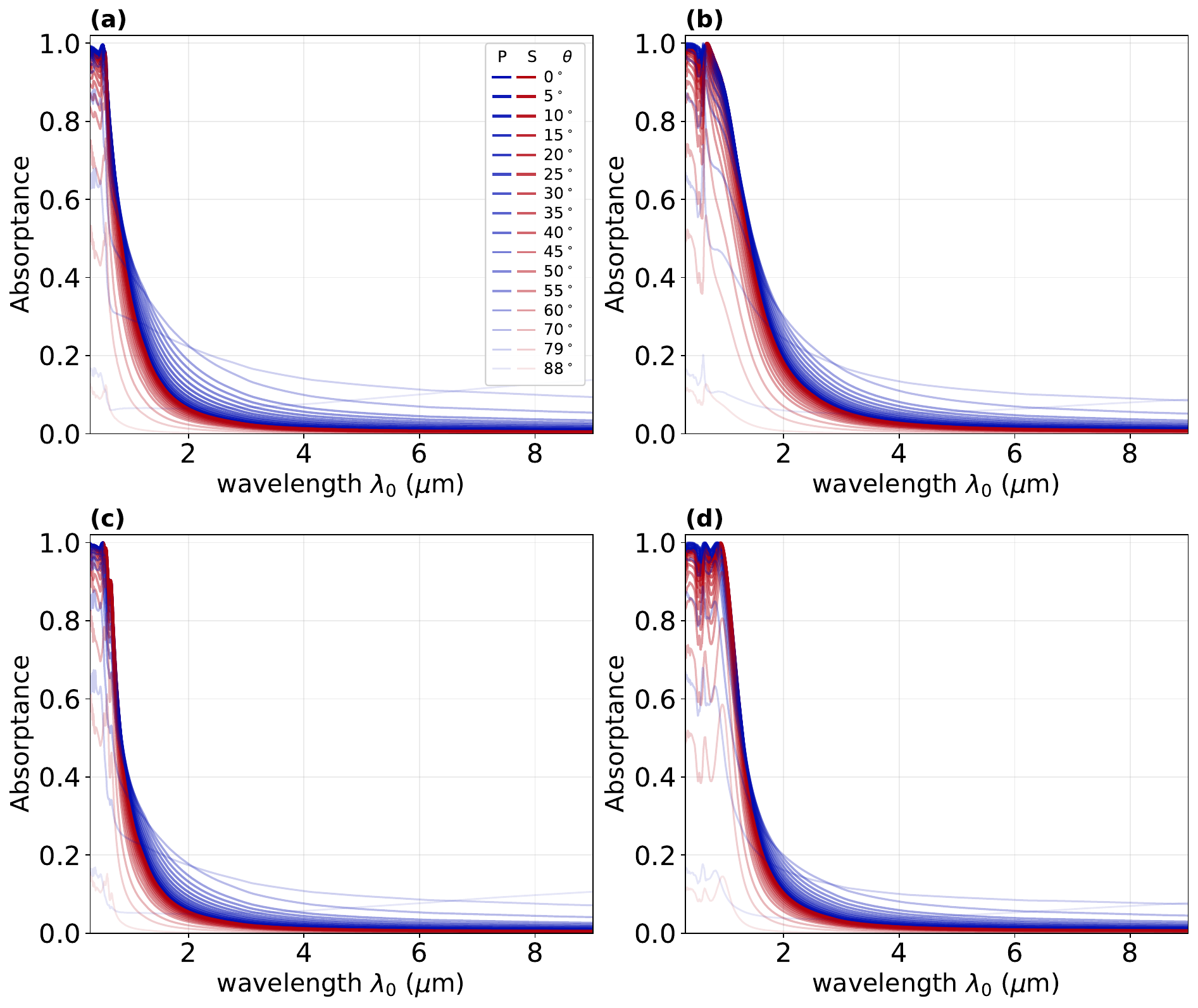}
\caption{Polarisation-resolved absorptance for (a) cuboid only, no spacer; (b) complete unit cell, no spacer; (c) cuboid only, 10~nm spacer; and (d) complete unit cell, 4~nm spacer. Blue/red curves show P/S polarisations, lightening from $0^\circ$ to $88^\circ$.}\label{fig:S4}
\end{figure}

The hemispherical estimate uses P- and S-polarised spectra over the full simulated angular and spectral ranges (Fig.~\ref{fig:S4}). P-polarised absorptance persists to larger incidence angles than S-polarised absorptance. The two are averaged before integration with the projected solid-angle weight.

\clearpage

\section{Azimuthal dependence}
\begin{figure}[htbp]
\centering\includegraphics[width=0.6\linewidth]{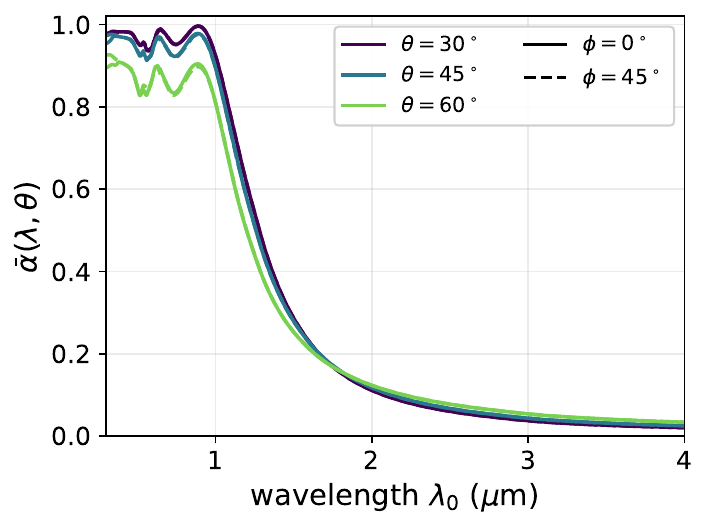}
\caption{Polarisation-averaged absorptance of the complete cell with a 4~nm spacer at $\phi=0^\circ$ (solid) and $45^\circ$ (dashed), for $\theta=30^\circ$, $45^\circ$ and $60^\circ$. Maximum pointwise solar-band differences are $0.4\%$, $1.9\%$ and $3.0\%$, respectively; the band average differs by at most $0.06\%$.}\label{fig:S5}
\end{figure}

The complete unit cell with a 4~nm spacer was compared at $\phi=0^\circ$ and $45^\circ$ for three polar angles (Fig.~\ref{fig:S5}). The band-averaged solar absorptance over $0.3$--$1.5~\mu$m differs by at most $0.06\%$. This is a sampled comparison, not a complete azimuthal bound on the Planck-weighted emission spectrum.

\clearpage

\section{Ideal-step spectral reference}
\begin{figure}[!htbp]
    \centering
    \includegraphics[width=1.0\linewidth]{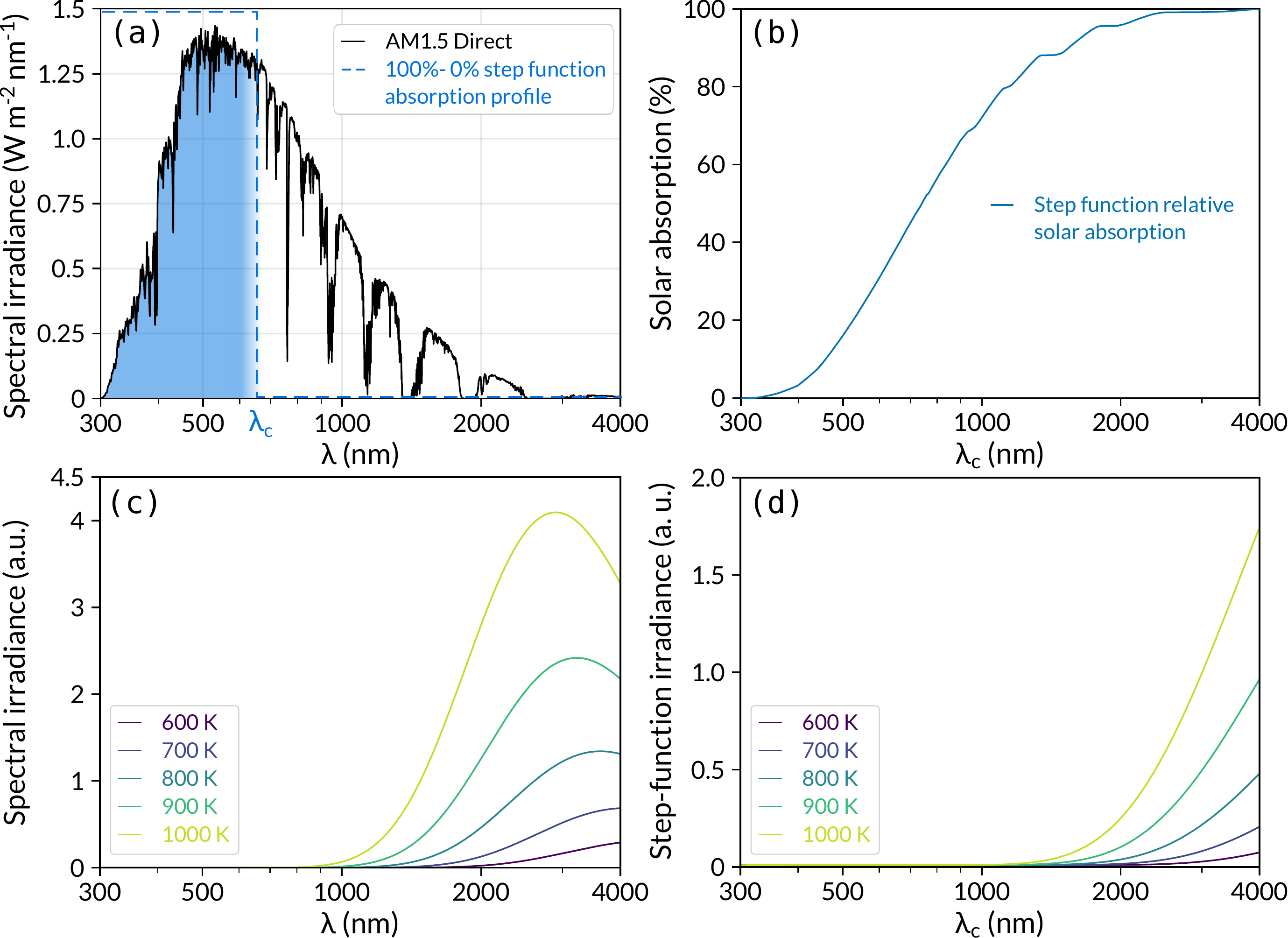}
    \caption{(a) AM1.5D solar spectrum and ideal step-absorber profile with cutoff wavelength $\lambda_c$. (b) Fraction of the AM1.5D spectrum absorbed below $\lambda_c$. (c) Blackbody spectral radiance at representative temperatures. (d) Cumulative emitted power of the ideal step absorber as a function of $\lambda_c$.}
    \label{fig:S6}
\end{figure}

The ideal-step profile in Fig.~\ref{fig:S6} illustrates how increasing the cutoff first adds useful AM1.5D power and then increases radiative loss as the Planck spectrum overlaps the high-absorptance band. The two-level profile and full radiative balance in the main text determine the reference cutoff; Fig.~\ref{fig:S6} alone does not set an optimum.

\clearpage

\section{Single-resonator field profiles}
\begin{figure}[!htbp]
    \centering
    \includegraphics[width=0.9\linewidth]{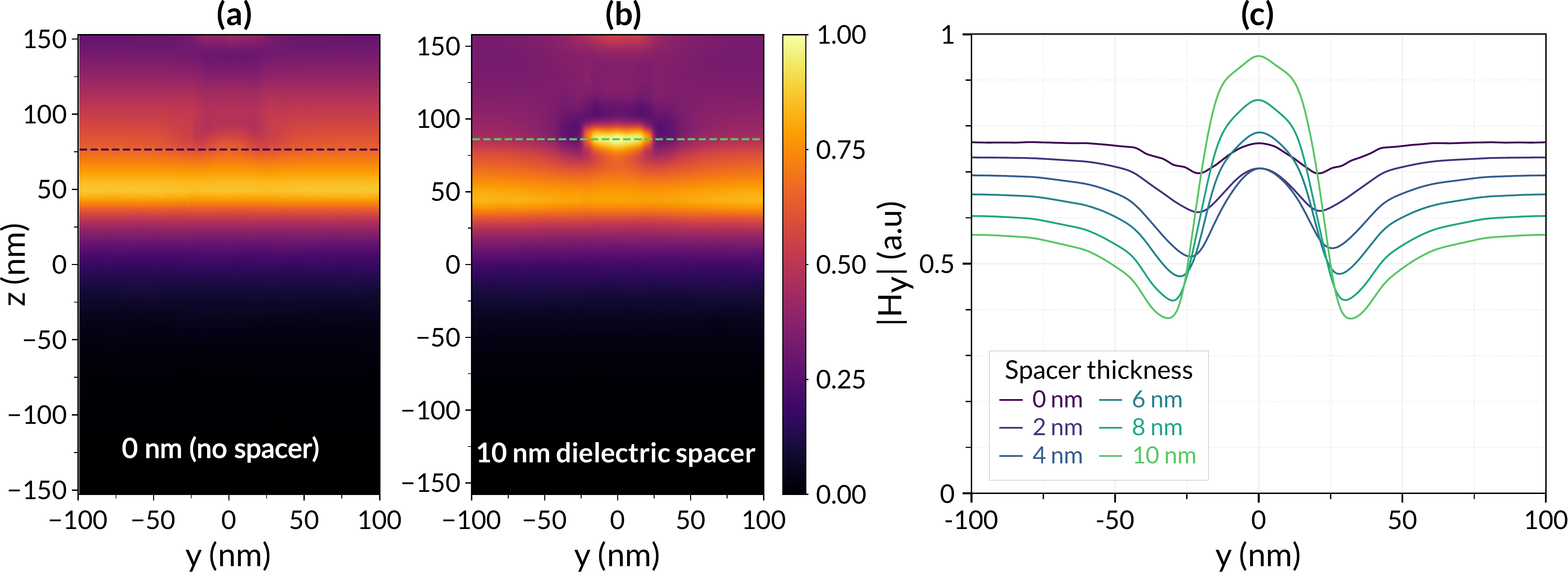}
    \caption{Magnetic field $|H_y|/H_\mathrm{i}$ in the $x$--$z$ plane through the cuboid ($y=0$) for (a) unit cell B and (b) unit cell C at $\lambda=633$~nm, and (d), (e) in the $y$--$z$ plane through the cuboid ($x=0$). (c) Corresponding $H_y$ line profiles beneath the gold resonator for spacer thicknesses from 0 to 10~nm, along $x$ (parallel to the incident electric field), and (f) along $y$ (normal to it), through the cell centre 1~nm below the resonator base (dashed lines in the maps; dotted lines: faces of the cuboid). All fields are normalised by the amplitude of the normally incident magnetic field, $H_\mathrm{i}$.}
    \label{fig:S7}
\end{figure}

At 633~nm, the spacer redistributes the field beneath the gold cuboid, as shown in Fig.~\ref{fig:S7}. The maps show the $x$--$z$ [Fig.~\ref{fig:S7}(a,b)] and the $y$--$z$ [Fig.~\ref{fig:S7}(d,e)] planes through the cuboid, and the profiles run through the cell centre along the incident electric field [Fig.~\ref{fig:S7}(c)] and normal to it [Fig.~\ref{fig:S7}(f)]. 1~nm below the base, $|H_y|$ at the centre of the cuboid becomes strongly localised and rises from $1.2\,H_\mathrm{i}$ without spacer to $3.1\,H_\mathrm{i}$ with the 10~nm spacer, while at the cell edge it changes from $1.3$ to $1.0\,H_\mathrm{i}$.

\clearpage

\section{Time-domain energy flow}
\label{sec:SI-td}
The frequency-domain calculations of the main text give the time-averaged power absorbed in each material delivered by monochromatic plane waves. Here, using the Ansys Lumerical Suite FDTD solver, we follow a short pulse impinging on the same four designs (unit cells B and C of the main text (cuboid only, without spacer and with a \SI{10}{nm} spacer) and the complete unit cells without spacer and with a \SI{4}{nm} spacer) to resolve how this energy reaches the carbon and partition the energy flow among the resonator layer, the carbon film and the gold reflector at every instant. 
Gold and carbon are modelled by passive multi-coefficient fits to the same optical data as in the frequency-domain calculations~\cite{babarWeaver2015,hagemannOpticalConstantsFar1975}. 

With Bloch boundary conditions, the solver propagates complex fields driven by a Gaussian-enveloped planewave $g(t)\,e^{-i\omega_0 t}$, with a \SI{468.4}{THz} carrier and a \SI{2.5}{fs} intensity full width at half maximum (FWHM; spectral FWHM 0.54--0.79~$\mu$m). The mesh step is \SI{1}{nm} along $z$ and \SI{2.5}{nm} laterally. Time monitors record all field components on $81\times81$ nodes per cell every \SI{0.244}{fs}, at an exterior plane \SI{300}{nm} above the resonator tops and at planes 1, 2 and \SI{3}{nm} inside the carbon film from both sides. 

At normal incidence, every FDTD update coefficient is real, so by linearity $\mathrm{Re}[e^{-i\varphi}\mathbf{E}]$ is the physical field of the real pulse $g(t)\cos(\omega_0 t+\varphi)$ for any carrier--envelope phase $\varphi$. The power crossing a plane $z$ towards the reflector, per unit cell of area $A_c$ and with $z$ directed away from the reflector,
\begin{equation}
P(z,t)=-\frac{1}{2}\int_{A_c}\mathrm{Re}\bigl[\mathbf{E}\times\mathbf{H}^{*}\bigr]_z\,\mathrm{d}^2r_\parallel ,
\label{eq:td-flux}
\end{equation}
is therefore the instantaneous Poynting flux averaged over $\varphi$. 
Poynting's theorem reads $-\nabla\cdot\mathbf{s}=\partial_t u+q$ with stored-energy and dissipation densities $u,q\geq0$. Integrated over the slab of the unit cell, where the periodic side walls cancel, for each region $X$ between two planes $z_1<z_2$,  it gives 
\begin{equation}
P_X(t)=P(z_2,t)-P(z_1,t)=\frac{\mathrm{d}U_X}{\mathrm{d}t}+Q_X(t),
\label{eq:td-balance}
\end{equation}
so a region power is defined as its dissipation rate plus the rate of change of its stored energy. The carbon power $P_\mathrm{C}=P(z_\mathrm{C},t)-P(0,t)$, between the upper face $z_\mathrm{C}$ and the lower face $z=0$ of the carbon film, and the reflector power $P_\mathrm{refl}=P(0,t)$ are both obtained by quadratic extrapolation from the three planes inside the carbon. At the exterior plane above the structure, only the zeroth diffraction order propagates, and its incident and reflected waves follow exactly from the cell-averaged fields (denoted by a bar), $\mathbf{E}_{\downarrow,\uparrow}=\tfrac{1}{2}\bigl[\bar{\mathbf{E}}\pm Z_0\hat{\mathbf{z}}\times\bar{\mathbf{H}}\bigr]_\perp$, with powers $P_{\downarrow,\uparrow}=A_c|\mathbf{E}_{\downarrow,\uparrow}|^2/2Z_0$. 
Referred to the resonator tops by the vacuum transit time, these define the resonator-layer power $P_\mathrm{res}=P_\downarrow-P_\uparrow-P(z_\mathrm{C},t)$, whose dissipation is that of the gold particles. By orthogonality of the Floquet harmonics, every plane power splits exactly into
\begin{equation}
P=P_\mathrm{u}+P_\mathrm{nf},\qquad
P_\mathrm{u}=-\frac{A_c}{2}\,\mathrm{Re}\bigl[\bar{\mathbf{E}}\times\bar{\mathbf{H}}^{*}\bigr]_z ,
\label{eq:td-split}
\end{equation}
where $P_\mathrm{u}$ is carried by the laterally uniform zeroth-order field and $P_\mathrm{nf}$ by the evanescent near field of the resonators. The orders do not couple inside the laterally homogeneous carbon, so $P_\mathrm{C,u}$ and $P_\mathrm{C,nf}$ are separately balanced channels (Eq.~\ref{eq:td-balance}) of energy delivery to the film. The energy delivered to region $X$, $\mathcal{E}_X(t)=\int_{-\infty}^{t}P_X\,\mathrm{d}t'$, gives its absorptance $A_X=\mathcal{E}_X(\infty)/\mathcal{E}_\mathrm{inc}$, with $\mathcal{E}_\mathrm{inc}=\int P_\downarrow\,\mathrm{d}t$. Since $Q_X\geq0$ and $U_X$ vanishes before and after the pulse, Eq.~\ref{eq:td-balance} also gives an exact lower bound, $U_X(t)\geq \mathcal{E}_X(t)-\mathcal{E}_X(\infty)$. For the resonator layer, subtracting the transit energy of the pulse through the same layers without particles estimates the energy held by the particles.

\begin{figure}[!htbp]
    \centering
    \includegraphics[width=\linewidth]{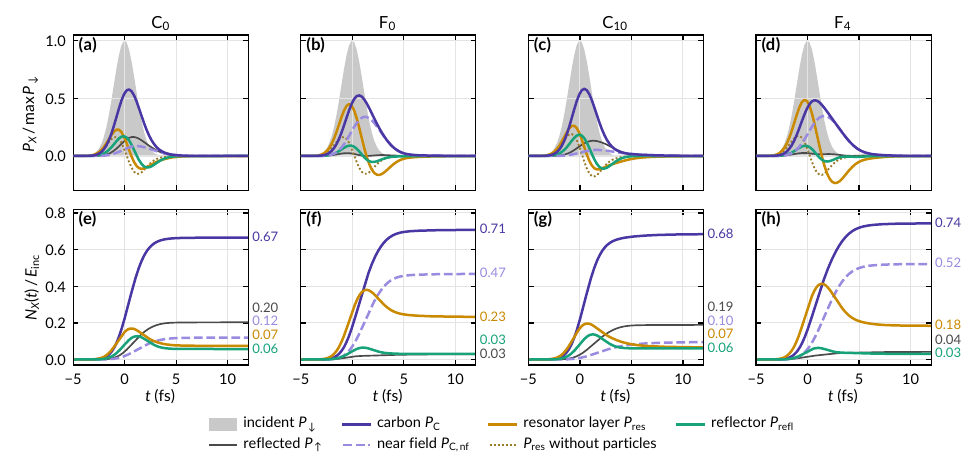}
    \caption{Region-resolved energy flow for the incident pulse. (a--d) Power entering the carbon film ($P_\mathrm{C}$, solid; near-field part $P_\mathrm{C,nf}$, dashed), the resonator layer ($P_\mathrm{res}$; dotted, the same layers without particles) and the reflector ($P_\mathrm{refl}$), with the incident (grey fill) and reflected powers, for (a) cuboid only, no spacer; (b) complete unit cell, no spacer; (c) cuboid only, 10~nm spacer; and (d) complete unit cell, 4~nm spacer, normalised to the peak incident power; $t=0$ is the incident peak at the resonator tops. (e--h) Delivered energies $\mathcal{E}_X(t)$ and reflected energy as fractions of $\mathcal{E}_\mathrm{inc}$, with their final values. Interfaces are extrapolated to their nominal positions at $\Delta z=\SI{1}{nm}$.}
    \label{fig:S8}
\end{figure}

Figure~\ref{fig:S8} shows the same sequence in all four designs. The resonator layer charges first as $P_\mathrm{res}$ rises ahead of the incident peak and turns negative once the pulse has passed, as the layer releases energy into the carbon and into the reflected wave. The carbon film receives energy throughout and never returns it, since $P_\mathrm{C}$ peaks shortly after the incident peak and stays positive. Recording the 3D field in the film shows that it stores at most 0.06--0.09~$\mathcal{E}_\mathrm{inc}$, so $P_\mathrm{C}$ is effectively the heating rate of the carbon. In the cuboid-only cells, much of the swing of $P_\mathrm{res}$ follows the particle-free reference (dotted), which shows the pulse crossing the 70~nm layer. Only the complete cells charge the layer well beyond this transit, and they release it in the deeper, later negative lobe of $P_\mathrm{res}$ [Fig.~\ref{fig:S8}(b,d)].

\begin{figure}[!htbp]
    \centering
    \includegraphics[width=\linewidth]{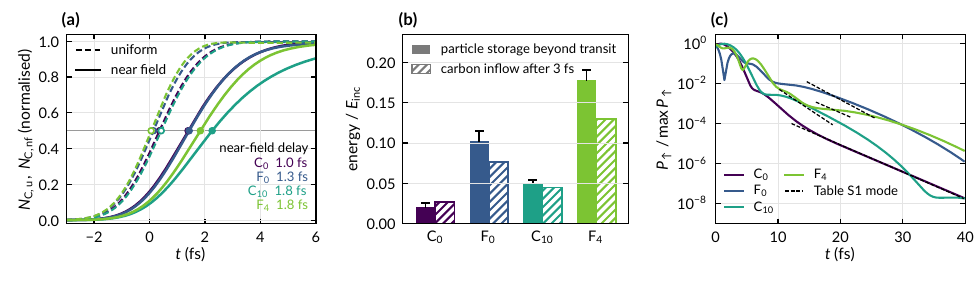}
    \caption{Timing of energy and power flow. (a) Progress of the uniform (dashed) and near-field (solid) carbon inflow, each normalised to its final value; circles mark the 50\% points, whose separation is the near-field delay: 1.0 and 1.3~fs for the cuboid-only and the complete cell without spacer, 1.8~fs for both cells with a spacer. (b) Reflected power referred to the resonator tops, relative to the peak incident power (grey), logarithmic scale.}
    \label{fig:S9}
\end{figure}

The dipole bars suppress the reflected energy, and most of the energy that is no longer reflected ends in their own gold: the carbon total rises much less than the absorption in the particles, as the frequency-domain absorptances also show. The bars change how the carbon film absorbs rather than how much it absorbs [Fig.~\ref{fig:S8}(e--h)]. Within the carbon, the bars exchange one channel for the other. The uniform field delivers less than half as much energy to the film as in the cuboid-only cells, and the near field becomes the main route into the carbon: $\mathcal{E}_\mathrm{C,nf}$ ends at about two-thirds of $\mathcal{E}_\mathrm{C}$ [dashed against solid end values in Fig.~\ref{fig:S8}(f,h)], against less than a fifth without bars. The near field delivers its energy later than the uniform field [Fig.~\ref{fig:S9}(a)]. The uniform part follows the incident pulse in all four designs, whereas the near-field part lags by 1.0~fs (cuboid only) and 1.3~fs (complete cell) without a spacer and by 1.8~fs with it. In the complete cells, the energy that the bars hold beyond transit [Fig.~\ref{fig:S8}(b,d)] reaches the carbon through this delayed near-field channel. Their long-axis modes are strongly damped (field lifetimes of about 1~fs, $Q\approx1.2$--1.5), so the bars act as low-$Q$ antennas that collect light over a wide band and pass it to the carbon within about one optical cycle.

The spacer changes the timing more than the final partition [Fig.~\ref{fig:S8}(c,d) against (a,b)]. It raises the energy held by the particles beyond transit, most of all in the complete cell, delays the near-field delivery and so increases the energy that enters the carbon after the incident pulse has passed [Fig.~\ref{fig:S9}(a)]; in the complete cells almost all of this late energy arrives through the near field. The partition itself changes at longer wavelengths, where the spacer red-shifts the bar long-axis mode from 0.67 to 1.05~$\mu$m.

Since the FDTD model is linear and time-invariant, the response to any pulse within the band of the incident one follows from the recorded fields up to the accuracy of the material fits in the target band, by Fourier transforming every record, multiplying it by the ratio of the new pulse spectrum to the incident one and transforming back. For a reference pulse centred at 1.05~$\mu$m (\SI{37}{THz} FWHM), obtained by exact spectral re-weighting of the same records, where the incident pulse still carries 5\% of its peak spectral power, the complete cells absorb most of the incident energy, nearly all of it in the carbon and mostly through the near field, and the spacer raises both their carbon share and the energy their particles hold, whereas the cuboid-only cells reflect much like the bare carbon film on gold. This yields a temporal counterpart of the near-infrared extension of the absorption band in the main text. The gold and carbon fits extend over this band, and for this pulse the absorption of every cell agrees within about 0.02 of the incident energy with the frequency-domain calculation from the tabulated optical data.

Figure~\ref{fig:S9}(b) shows the reflected power on the scale of Fig.~\ref{fig:S8}. In the complete cells, it dips at instants at which the reflected field nearly vanishes, and these dips show how the bars suppress the reflection. Subtracting the reflected field of the cuboid-only cell with the same spacer leaves the field that the bars add. Throughout the pulse, this field is nearly opposite in phase to the reflection of the cuboid-only cell, so that the complete cells reflect several times less power, and the cancellation is best wherever their relative phase passes exactly through $\pi$. Without a spacer, these two fields, reflected from the cuboid and dipole bars separately, also have nearly equal amplitudes, so the reflected power collapses about 1.5~fs after the incident peak. With the spacer, the field of the bars is a fifth to a third stronger: the first passage, during the pulse, leaves only a shallow minimum, and the deep dip comes at the second, at about 4~fs, after the pulse, when both fields have decayed. The reflection that follows is the emission of the energy the bars still hold. All four responses are essentially over within about 10~fs: by then the reflected power has fallen below $10^{-3}$ of the incident peak.

Because the response is linear, every time-integrated quantity is fixed by the spectral absorptances, and under stationary solar illumination the absorbed power depends only on $\alpha(\lambda)$. The transients resolve instead the route of the absorbed energy: the bars pass light collected in broad, strongly damped modes to the carbon through the near field, and the spacer increases transient storage, delays this delivery and red-shifts the bar resonance into the near infrared.

\clearpage

\section{Angle-resolved maps}
\begin{figure}[!htbp]
    \centering
    \includegraphics[width=\linewidth]{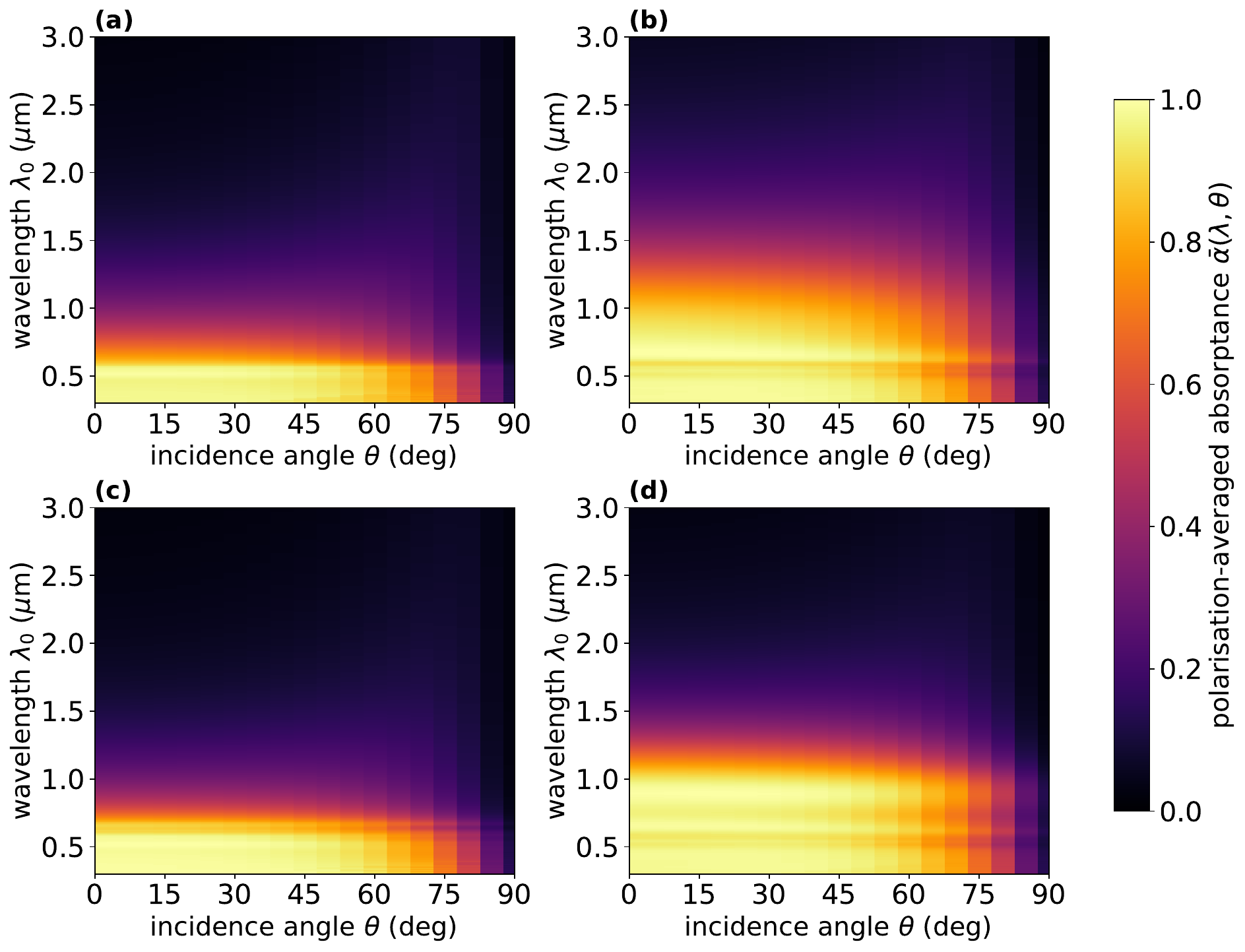}
    \caption{Polarisation-averaged absorptance $\bar{\alpha}(\lambda,\theta)=\tfrac{1}{2}[\alpha_P(\lambda,\theta)+\alpha_S(\lambda,\theta)]$ as a function of wavelength and polar incidence angle for the four metasurface configurations: (a) cuboid only, no spacer; (b) complete unit cell (cuboid and dipole bars), no spacer; (c) cuboid only, 10~nm idealised lossless dielectric spacer ($n=1.45$); (d) complete unit cell, 4~nm idealised lossless dielectric spacer ($n=1.45$).}
    \label{fig:S10}
\end{figure}

The maps in Fig.~\ref{fig:S10} show a broad solar-band response up to large polar angles and suppressed infrared absorptance over the plotted range. Polarisation-resolved data over 0.3--9~$\mu$m appear in Fig.~\ref{fig:S4}. The maps show the absorptance itself without the geometric factor $\cos\theta$ of the projected emission and the $\sin\theta$ of the solid-angle element; they enter in the hemispherical emissivity of the main text (Eq.~11).
The main text reports the hemispherically weighted results based on the polarisation-averaged angular spectra shown here.

\clearpage
\section{Mesh convergence and rounded resonator corners}
\label{sec:SI-fillets}
\begin{figure}[!htbp]
    \centering
    \includegraphics[width=\linewidth]{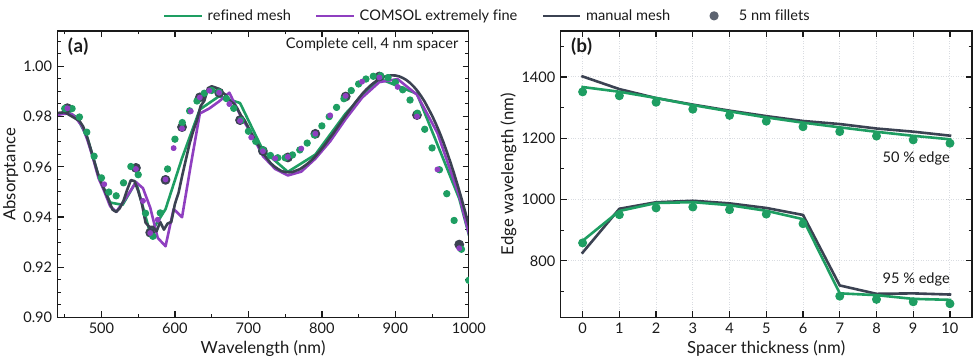}
    \caption{Mesh convergence and rounded corners of the complete unit cell at normal incidence. (a) Absorptance of the cell with the 4~nm spacer on the refined mesh (green), COMSOL's extremely fine mesh (violet) and the manual mesh used in the main text (dark grey), with sharp corners as lines and rounded corners as markers. (b) 95\% and 50\% edges of the spacer sweep on the manual mesh (Fig.~8 of the main text) and on the refined mesh, with sharp and rounded corners. Refined mesh entails 4~nm elements on the resonator surfaces; growth rate 1.25 in carbon and air, and 1.05.}
    \label{fig:S11}
\end{figure}

The results of the main text use manual meshing with the maximum element size $h_{\max}=\lambda_0/(N n_{\mathrm{mat}})$ with $N=6$ in each material and a growth rate of 1.05, and the resonators have sharp in-plane corners. A convergence check by raising $N$ to 10 in carbon and air changes the absorptance of the complete unit cell with the 4~nm spacer by at most 0.004. The complete cells without and with the 4~nm spacer were also computed on two different meshes with finer gold elements: a refined manual mesh with $h_{\max}=$4~nm on the resonator surfaces, and COMSOL's Extremely Fine auto-mesh setting. To rule out the effect of singularity in metallic sharp corners, every above-mentioned mesh was also used with the resonator corners rounded by 5~nm fillets (0.6\% less resonator gold).
The spectra nearly coincide, as Fig.~\ref{fig:S11}(a) shows for the cell with the 4~nm spacer, up to a slight shift as a result of rounding the corners, which changes the shape of the resonator particles.
Across the spacer sweep, rounding the corners moves the 95\% and 50\% edges by 7--16~nm to shorter wavelengths [Fig.~\ref{fig:S11}(b)].

\clearpage
\FloatBarrier

\end{document}